\documentclass[reqno,tbtags,intlimits,a4paper,oneside,12pt]{amsart}
\usepackage[cp1251]{inputenc}%
\usepackage{amssymb,upref,calc,url} 
\usepackage[mathscr]{eucal}
\usepackage{amscd}
\usepackage[in,myheadings]{fullpage}
\newtheorem{theorem}{Theorem}
\newtheorem{cor}[theorem]{Corollary}
\newtheorem{lemma}[theorem]{Lemma}
\theoremstyle{definition}
\newtheorem{definition}[theorem]{Definition}
\newtheorem{example}[theorem]{Example}
\newtheorem{cexample}[theorem]{The classical Example}

\newtheorem{rem}[theorem]{Remark}
\newtheorem{rems}[theorem]{Remarks}

\allowdisplaybreaks[4] 

\begin{document}

\title[V. M. Buchstaber, E. Yu. Bunkova.
Elliptic formal group laws.] {Elliptic formal
group laws, \\ integral Hirzebruch genera and Krichever genera.}
\author{V. M. Buchstaber, E. Yu. Bunkova}
\address{Steklov Mathematical Institute, Russian Academy of sciences,
Moscow, Russia.} \email{buchstab@mi.ras.ru} \email{bunkova@mi.ras.ru}

\maketitle 

\section*{\bf Introduction.}

The theory of formal group laws plays an important role in many directions of modern mathematics. In the classical papers and wellknown textbooks (see \cite{ref17}, \cite{ref28}, \cite{ref29}, \cite{ref30})
we can find fundamental results
concerning the group structure on elliptic curves.
This results lead to remarkable formal group laws.
We consider an elliptic curve as an irreducible non-singular projective
algebraic curve of genus $1$ furnished with a point $0$, the zero of the
group law. Any such curve has a plane cubic model of
the form
\[
y^2 + \mu_1 x y + \mu_3 y = x^3 + \mu_2 x^2 + \mu_4 x  + \mu_6.
\]
In Tate coordinates the geometrical addition laws on this curves
correspond to the general formal group law over the ring
$\mathbb{Z}[\mu_1, \mu_2, \mu_3, \mu_4, \mu_6]$. We study the
structure of this law and the differential equation that determines its exponent. 
We describe a 5-parametric family of Hirzebruch genera with integer values on
stably complex manifolds. We introduce the general Krichever genus, which is given by a generalized
Baker-Akhiezer function.
 This function has many of the fundamental properties of the Baker-Akhiezer function, but unlike it, it is not meromorphic,
because it can have two branch points in the parallelogram of periods.

In section 1 we give necessary facts on Hurwitz series, formal group laws and Hirzebruch genera and formulate the problems connecting this three fundamental objects. The main part of the paper contains the results on this problems in the cases when this objects are defined by the elliptic curves.

\tableofcontents

\section{\bf Main notions and problems.}

Let $A$ be a commutative associative evenly graded torsion-free ring $A = \sum_{k \geq 0} A_{- 2 k}$
with identity element $1 \in A_0$.

\subsection{Graded Hurwitz series.}
A {\it Hurwitz series} over $A$ is a formal power series in the form
 \[ \varphi(u)=\sum_{k\geqslant 0}\varphi_k \frac{u^k}{k!}
\;\in\; A\otimes \mathbb{Q}[[u]]
\]
with $\varphi_k \in A$ for all $k = 0, 1, 2, 3,\ldots$ (see \cite{ref20}).
Set $\deg u = 2$. The series $\varphi(u)$ has the degree $2 l$ if
$\deg \varphi_k + 2 k = 2 l$ for any $k$. Let us denote the set of
Hurwitz series over $A$ by $H A [[u]]$ and the set of graded Hurwitz
series over $A$ by $H^{gr} A [[u]]=\sum H_{2l}^{gr}A [[u]]$.

{\bf Properties of Hurwitz series:}

1) Hurwitz series over $A$ form a commutative associative ring $H A
[[u]]$ with respect to the usual addition and multiplication of
series.

2) The units of the ring $H A [[u]]$ are the elements $\psi(u)$ such
that $\psi(0)$ is a unit of $A$.

3) This ring is closed under differentiation and integration with
respect to $u$.

4) If $\varphi(u) \in H A [[u]]$, $\varphi(0) = 0$, $\psi(u) \in H B
[[u]]$, then $\psi(\varphi(u)) \in H (A \otimes B) [[u]]$.

If $\varphi(u) \in H A [[u]]$, $\varphi(0) = 0$, $\varphi'(0)$ is a
unit of $A$, then $\varphi^{-1}(u) \in H A [[u]]$, where
$\varphi^{-1}(u)$ is defined by $\varphi(\varphi^{-1}(u)) = u$.

\subsection{The formal group law.}
A commutative one-dimensional {\it formal group law} over the ring $A$ is a formal
series
$
\; F(t_1, t_2) = t_1 + t_2 + \sum_{i,j} \alpha_{i,j} t_1^i t_2^j, \; \, \alpha_{i,j} \in A, \;
$
such that the following conditions hold:
\[
F(t, 0) = t, \quad F(t_1, t_2) = F(t_2, t_1),\quad F(t_1, F(t_2, t_3)) = F(F(t_1, t_2), t_3).
\]
For brevity we will use "formal group" \,as synonim to "formal group law".

A {\it homomorphism} of formal groups $h: F_1 \to F_2$ over a ring $A$
is a formal series $h(u) \in A[[u]]$ where $h(0)=0$ such that
$h(F_1(u,v)) = F_2(h(u),h(v))$. The homomorphism $h$ is an
{\it isomorphism} if $h'(0)$ is a unit in $A$ and a {\it strong isomorphism} if
$h'(0) = 1$.

For each formal group $F \in A[[t_1,t_2]]$ there exists a strong
isomorphism $f: L \to F$ over the ring $A \otimes \mathbb{Q}$, where
$L$ is the linear group $L(u,v) = u+v$. It is uniquely defined by the
condition
\begin{equation} \label{f}
f(u+v) = F(f(u),f(v)).
\end{equation}
The function $f(u)$ is called the {\it exponential} of the formal group
$F$. The function $g(t) = f^{-1}(t)$ is called the {\it logarithm} of the
formal group  $F$. From (1) we have
\begin{equation} \label{f'}
f'(u) = {\partial \over \partial t_2} F(f(u),t_2) |_{t_2 = 0}\,, \qquad {1
\over g'(t)} = {\partial \over \partial t_2} F(t,t_2) |_{t_2 = 0}\,.
\end{equation}
Thus, $g'(t)$ and $f'(g(t))\in A[[t]]$. Therefore $g(t)$ and $f(u)$ are Hurwitz
series. If $F$ is graded of degree $2$, (that is $\deg \alpha_{i,j} = - 2
(i+j-1)$), then $f(u) \in H_2^{gr} A[[u]]$ and $g(t) \in H_2^{gr} A[[t]]$.

Let $F_1$ and $F_2$ be formal groups over $A$. There exists a strong isomorphism $h: F_1 \to F_2$ over $A$ if and
only if $f_2\big(f_1^{-1}(t)\big) \in A[[t]]$. Notice that if $h:
F_1 \to F_2$ a strong isomorphism, and $f_1$ is the exponential of
$F_1$, then $h(f_1(u+v)) = F_2(h(f_1(u)),h(f_1(v)))$, that is $h
\circ f_1$ is the exponential of $F_2$.
Thus there exists a strong
isomorphism $\; h: L \to F_2 \;$ over $\; A$ \\ if and
only if the exponential $f(u)$ is a series over $A$.

Let $f(u) \in H_2^{gr} A[[u]]$, $f(0) = 0$, $f'(0) = 1$, and $F(t_1,t_2) = f(f^{-1}(t_1) +
f^{-1}(t_2))$. We have $F(t_1,t_2) = t_1 + t_2 + \sum_{n \geq 1} \sum_{i+j =
n+1} \alpha_{i,j} t_1^i t_2^j \in H_2^{gr} A[[t_1, t_2]]$. The
coefficients $\alpha_{i,j}$ are polynomials of $f_i$, where $f(u) =
u + \sum_{i \geq 1} f_i u^{i+1}$. By the construction
$\alpha_{i,j} \in A_{-2(i+j-1)} \otimes \mathbb{Q}$. Denote by $A_f$
the subring in $A \otimes \mathbb{Q}$ generated by $\alpha_{i,j}$
and by $B_f$ the subring in $A \otimes \mathbb{Q}$ generated by
$f_i$. The ring $B_f$ is by the construction the smallest ring over
which there exists a strong isomorphism between $L$ and $F$. Thus we have
$A_f \subset B_f \subset A \otimes \mathbb{Q}$.

Thus, {\it the problem of description of graded formal groups over a
torsion free ring $A$ is equivalent to the problem of description of
such $f(u) \in H_2^{gr}A[[u]]$, that $F(t_1,t_2) \in A[[t_1, t_2]]$}. Here
arises the {\it problem of description of the graded rings
$A_f$ and $B_f$}.

\subsection{The universal formal group law.} \text{}

A formal group law $\mathcal{F}(t_1,t_2) = t_1 + t_2 +
\sum a_{i,j} t_1^i t_2^j$ over the ring $\mathcal{A}$ is the {\it universal formal group law} if
for any formal group  $F(t_1,t_2)$ over any ring  $A$ there exists an
unique ring homomorphism $r: \mathcal{A} \to A$ such that $F(t_1,t_2) =
t_1 + t_2 + \sum r(a_{i,j}) t_1^i t_2^j$.

Thus, the {\it problem of description of formal groups over $A$ can
be brought to the problem of description of ring homomorphisms}
$\mathcal{A} \to A$.

Consider the graded ring $\mathcal{U} = \mathbb{Z}[\beta_{i,j}],\;
i>0,\; j>0,\; \deg \beta_{i,j} = - 2(i+j-1)$ and the series
$$\Phi(t_1,t_2) = t_1 + t_2 +\sum \beta_{i,j} t_1^i t_2^j.$$
Set $\Phi^l(t_1,t_2,t_3) = \Phi\big(\Phi(t_1,t_2),t_3\big)$ and $\Phi^r(t_1,t_2,t_3) =
\Phi\big(x,\Phi(y,z)\big)$. We obtain
\begin{align*}
\Phi^l(t_1,t_2,t_3) &= t_1 + t_2 + t_3 + \sum \beta^l_{i,j,k} t_1^i t_2^j t_3^k,\\
\Phi^r(t_1,t_2,t_3) &= t_1 + t_2 + t_3 + \sum \beta^r_{i,j,k} t_1^i t_2^j t_3^k
\end{align*}
where $\beta^l_{i,j,k}$ and $\beta^r_{i,j,k}$ are homogeneous
polynomials of $\beta_{i,j}$, $\deg \beta^l_{i,j,k} = \deg
\beta^r_{i,j,k} = \\ -2(i+j+k-1)$.

Let $J\subset \mathcal{U}$ be the ideal of associativity generated
by polynomials $\beta^l_{i,j,k}-\beta^r_{i,j,k}$. Consider the ring
$\mathcal{A}=\mathcal{U}/J$ and the canonical projection $\pi \colon
\mathcal{U} \to \mathcal{A}$. Denote by $\mathcal{F}(t_1,t_2)$ the
series $\pi[\Phi](t_1,t_2)=t_1+t_2+\sum \pi(\beta_{i,j}) t_1^i t_2^j.$

By the construction, the formal group  $\mathcal{F}(t_1,t_2)$ over
the ring $\mathcal{A}$ is the universal formal group. Therefore,
by Lazard's theorem (see \cite{ref18}) we obtain that
\[ \mathcal{A}=\mathbb{Z}[a_n],\; n=1,2,\ldots, \; \deg a_n=-2n. \]
Thus, the {\it problem of description of formal groups over the ring
$A$ is equivalent to the problem of description of ring
homomorphisms}\, $\mathbb{Z}[a_n] \to A$.

Consider the exponential $f_{\mathcal{U}}(u)=u+\sum b_n u^{n+1}\in
\mathcal{A}\otimes \mathbb{Q}[[u]]$ of the group $\mathcal{F}(t_1,t_2)$
and the ring $\mathcal{B} = \mathbb{Z}[b_n]\subset
\mathcal{A}\otimes \mathbb{Q}$.

Thus the  problem of description of formal groups over
torsion-free rings in terms of their exponentials can be presented in
the following way:

For a formal group  $F(t_1,t_2) = \sum \alpha_{i,j} t_1^i t_2^j$ over the
torsion-free ring $A$ the classifying group homomorphism
holds:
\[
\phi: \mathcal{A} \longrightarrow A.
\]
It is defined by the condition $\phi(\beta_{i,j}) = \alpha_{i,j}$.

Having $F(t_1,t_2) \in  A[[t_1,t_2]]$, we find $g(t)\subset
A \otimes \mathbb{Q}[[t]]$ from (\ref{f'}) and the exponential
$g^{-1}(u) = f(u) = u+\sum f_n u^{n+1}$. Because $f_{\mathcal{U}}(u)= u + \sum b_n u^{n+1}$
we get $\phi(b_n)=f_n$.

We obtain the commutative diagram:
\[
\begin{CD}
0 @ >>> J_F @ >>> \mathcal{A} @ >>> A_f @ >>> 0\\
@.@ VVV @ VVV @ VVV @. \\
0 @ >>> \widehat J_F @ >>> \mathcal{B} @ >>>  B_f @ >>> 0
\end{CD}
\]

{\it The problems of the description of the kernels $J_F,\; \widehat
J_F$ and the images $ A_f,\; B_f$ for $\mathcal{A},\;
\mathcal{B}$ arise}.

\begin{cexample}
Let $f(u) = {1 \over \mu} (1 - e^{- \mu u})\in H_2^{gr}A[[u]]$ where
$A=\mathbb{Z}[\mu]$, $\deg \mu = -2$. Then we have $g(t) = - {1 \over \mu}
\ln(1 - \mu t)\in H_2^{gr}A[[t]]$ and $F(t_1,t_2) = t_1 + t_2 - \mu t_1 t_2$ is
a formal group  over $\mathbb{Z}[\mu]$. Then $A_f =
A$ and $B_f$ is isomorhic to $\mathbb{Z}[b_1,\ldots, b_n,\ldots] / J$ such that
$A_f \rightarrow B_f: \mu \mapsto - 2 b_1$. We have $f_i
= (-1)^i {\mu^i \over (i+1)!}$, thus $J$ is generated by polynomials
$(i+1)! b_i - 2^i b_1^i$, $i = 2, 3,\ldots$
\end{cexample}

Denote by $\mathcal{A}^{(1)} = \sum \mathcal{A}_{-2n}^{(1)}$ the quotient
ring of the ring $\mathcal{A}$ modulo an ideal $I$, which is
generated by the elements, that can be represented as a product of
at least two elements of positive degree. Let $\widetilde{\pi}: \mathcal{A} \to
\mathcal{A}^{(1)}$ be the canonical projection. We obtain \\
$\widetilde{\pi}[f_{\mathcal{U}}^{-1}](t) = t - \sum b_n t^{n+1}$ and
\[
\widetilde{\pi}[\mathcal{F}](t_1,t_2) = t_1 + t_2 + \sum b_n ((t_1+t_2)^{n+1} - t_1^{n+1} - t_2^{n+1}).
\]
Thus the subgroup $\widetilde{\pi}(\mathcal{A}_{-2n})$ of the group
$\mathcal{A}_{-2n}^{(1)}$ is isomorphic to the group $\mathbb{Z}$ with \\
generators $\nu(n+1) \widetilde{\pi}(b_n)$, where
 $\nu(n) = \text{gcd} \begin{pmatrix} n \\ k \end{pmatrix}, k = 1, 2,\ldots,
n-1$.

There exist multiplicative generators $a_n^*,\; n=1,2,\ldots$ of the ring $\mathcal{A}=\mathbb{Z}[a_n]$, such that the embedding $\mathcal{A} \to \mathcal{B}$ is given by the formula $a_n^*=\nu(n+1)b_n^*$, where $b_n^*,\; n=1,2,\ldots$ are the multiplicative generators of the ring
$\mathcal{B} = \mathbb{Z}[b_n]$.

{\bf The choice of generators in $\mathcal{A}$} (see survey \cite{ref4}).

For any formal group $F(t_1,t_2)$ over $A$ one can define {\it the power system} (see \cite{ref3}): \\
$[t]_n \in A[[t]]$ for $n \in \mathbb{Z}$, where $[t]_n = n t + (t^2)$, $[t]_0 = 0$, $[t]_1 = t$,
and $[t]_n$ are defined recursively by the condition $[t]_n = F(t, [t]_{n-1})$, $n = 2,3,4,\dots$, and $n = 0, -1, -2, \dots$.
\\We will denote $\bar{t} = [t]_{-1}$. Over the torsion-free ring $A$ we have $[t]_k = f(k \, g(t))$.

Let $p$ be a prime number, $\mathbb{Z}_{(p)}$ - the ring of integer $p$-adic numbers. In the polynomial ring
$\mathcal{A}_{(p)} = \mathcal{A} \otimes \mathbb{Z}_{(p)}$ the multiplicative generators can be chosen in the following way: Set
\begin{equation} \label{generators}
{\partial \over \partial t_2} \mathcal{F}(t, t_2) |_{t_2 = 0} = \sum_n c_n t^{n},
\qquad
[t]_p = \sum_n \widehat{c}_n t^{n + 1},
\end{equation}
then $\mathcal{A}_{(p)} = \mathbb{Z}_{(p)}[\widehat{a}_n]$, where
$  \widehat{a}_n = c_n$ for $n \neq p^q - 1$ and $\widehat{a}_n = \widehat{c}_n$ for $n = p^q - 1$, $q = 1,2,3, \dots$.

For $p=2$ the generators can be also chosen in the following way: \\
Let $\gamma(t) = \overline{t}$. We have $\gamma(t) = - t + \sum_{i\geq 1} \gamma_i  t^{i+1}$. Then $\mathcal{A}_{(2)} = \mathbb{Z}_{(2)}[\widehat{a}_n]$, where $\widehat{a}_n = c_n$ for $n\neq 2^q -1$ and $\widehat{a}_n = \gamma_{2^q-1}$ for $n = 2^q -1$,  $q = 1, 2, ...$.

The multiplicative generators of $\mathcal{A}_{(p)}$ in the dimentions $-2(p^q - 1)$
can be chosen using the coefficients of the series $[t]_{1-p}$ instead of the coefficients of the series $[t]_p$.
In the case $p=2$ these are the coefficients of $\overline{t}$.
The proof follows from the identity $F([t]_p, [t]_{1-p})= t$ and the fact that $[t]_p = pt + (\text{series with coefficients of negative degree})$.

\begin{lemma}Let $f(u) \in H A [[u]]$, $f(0) = 0$, $f'(0) = 1$. Then $f(g(t_1)+ g(t_2)) \in A_{(p)}$.
\end{lemma}

{\bf Proof.} Consider the formal group $F(t_1, t_2) = f(g(t_1)+g(t_2)) \in A \otimes \mathbb{Q}[[t_1, t_2]]$ and the homomorphism $\mathcal{A} \to A \otimes \mathbb{Q}$, classifying this group. Using the choice of multiplicative generators of the ring $\mathcal{A}$ described above, we get that the image of the ring $\mathcal{A}$ lies in $A_{(p)}$.

\subsection{The formal group of geometric cobordisms.}

Consider the ring of complex cobordisms $\Omega_U$ of stably-complex manifolds (see \cite{ref19}, \cite{ref34}). According to Milnor-Novikov theorem we have $\Omega_U\simeq \mathbb{Z}[a_n],\; n=1,\ldots,\;
\deg a_n=-2n$.

Let $\eta \to \mathbb{C}P^n,\; n\leqslant\infty$ be the canonical complex line bundle over an $n$-dimentional complex projective space.

In the theory of complex cobordisms the isomorphism
$U^*(\mathbb{C}P^n)=\Omega_U[t]/(t^{n+1})$ takes place, where $t=c_1(\eta)$ is the first Chern class.
The formal group
\begin{equation} \label{Fcc}
\mathcal{F}_U(t_1,t_2) = t_1 + t_2 + \sum c_{i,j} t_1^i t_2^j, \quad \deg c_{i,j} = -2(i+j-1),
\end{equation}
of complex cobordisms is given by series (see \cite{ref9})
\[
c_1(\eta_1\otimes\eta_2)=\mathcal{F}_U(t_1,t_2)\in U^*
(\mathbb{C}P^\infty\times \mathbb{C}P^\infty)\cong
\Omega_U[[t_1,t_2]].
\]
The logarithm of this group (the series of
A.S.~Mishchenko) takes the form
\begin{equation}
g(t) = \sum\limits_{n\geqslant0}[\mathbb{C}P^n]\frac{t^{n+1}}{n+1}.
\end{equation}

In the work \cite{ref9} the Adams-Novikov operators $\Psi^k_U$ in the theory of complex cobordisms were introduced in therms of formal groups $\mathcal{F}_U(t_1,t_2)$. This operators are defined by the formulas $\Psi^k_U t = {1 \over k} [t]_k$, $k \ne 0$, and \; $\Psi^0_U t = g(t)$. \\ Notice that $\Psi^k_U t \in U^*(\mathbb{C}P^{\infty}) \otimes \mathbb{Z}[{1 \over k}]$, $k \ne 0$, and $\Psi^0_U t \in U^*(\mathbb{C}P^{\infty}) \otimes \mathbb{Q}$.

S.P.~Novikov showed the important role of the operators $\Psi^k_U$ for the Adams-Novikov spectral sequence and for the description of cobordism classes of manifolds with a group action in terms of fixed points of this action.

D.~Quillen published in \cite{ref11} the fundamental observation, that the homomorphism
 $\varphi \colon \mathcal{A} \to \Omega_U$, classifying the formal group $\mathcal{F}_U(t_1,t_2)$, is an isomorphism, thus the formal group of complex cobordisms was identified with the universal formal group. In this work on the basis of the results of the algebraic theory of formal groups an effective description of the important theory of Brown-Peterson cohomology was obtained. The applications of the theory of formal groups formed the background of a powerful direction in algebraic topology (for the foundations of this direction see the survey \cite{ref4}).

\eject

\subsection{Hirzebruch genera and formal groups.}

Let
\[
f(u)=u + \sum_{k \geq 1} f_k u^{k+1}, \quad \text{where} \quad f_k \in A\otimes \mathbb{Q}.
\]
The formal series
\[
\prod\limits_{i=1}^n \frac{u_i}{f(u_i)}
\]
is invariant with respect to the permutation of the variables $u_1,\ldots,u_n$, and therefore
it can be presented in the form
$L_f(\sigma_1,\ldots,\sigma_n)$, where $\sigma_k$ is the $k$-th elementary symmetric polynomial of $u_1, \ldots, u_n$.

{\it The Hirzebruch genus $L_f(M^{2n})$ of a stably complex manifold}
$M^{2n}$ is the value of the cohomology class
$L_f(c_1,\ldots,c_n)$ on the fundamental cycle $\langle
M^{2n}\rangle$ of the manifold $M^{2n}$, where $c_k$ is the
 $k$-th Chern class of the tangent bundle of the manifold $M^{2n}$.
The foundations of the theory and the applications of the Hirzebruch genera were laid in the work \cite{ref27}.

Each Hirzebruch genus $L_f$ defines a ring homomorphism
\[
L_f \colon \Omega_U \to A\otimes \mathbb{Q}\,
\]
and for any ring homomorphism $\varphi \colon \Omega_U \to A \otimes \mathbb{Q}$
there exists a series $f(u) \in A\otimes \mathbb{Q}[[u]],$
$f(0)=0,\; f'(0)=1$ such that $\varphi=L_f$.

The Hirzebruch genus $L_f$ is called {\it $A$-integer} if $L_f(M^{2n}) \in A_{-2n}$ for any stably complex manifold $M^{2n}$.

Identifying the formal group of complex cobordisms with the universal formal group, we obtain (see details in \cite{ref3}):

The correspondence of the formal group $F(t_1,t_2)$ over the ring $A$ to
its exponential $f(u)$~$\in$~$HA[[u]]$
gives a one-one mapping of the formal groups $F(t_1,t_2)$ over the ring $A$ and the Hirzebruch genera $L_f$, taking values in $A$.

Thus,
{\it the problem of description of series $f(u)\in HA[[u]]$ such that} $F(t_1,t_2) \in
A[[t_1,t_2]]$ (see section 1.2), {\it is equivalent to the problem of description of $A$-integer
Hirzebruch genera}.

The importance and the celebrity of this problem (the famous Atiyah–Singer theorem and its developments) is connected with the Hirzebruch genera,
the values of which on manifolds is equal to indices of fundamental differential operators on this manifolds.

S.P.~Novikov in \cite{ref10} showed that for numerical Hirzebruch genera the formula holds:
\[ \sum L_f([\mathbb{C}P^n])\frac{t^{n+1}}{n+1}=g(t). \]
Thus if we know the formal group we obtain the values of $L_f([\mathbb{C}P^n])$ using the formula (\ref{f'}).
In the general case it is shown in \cite{ref7} that the identity map $\Omega_U \to \Omega_U$ gives the Chern-Dold character
\[
ch_U: U^*(X) \to H^*(X; \Omega_U(\mathbb{Z})) \subset  H^*(X; \Omega_U \otimes \mathbb{Q}),
\]
where $H^*(\cdot \,; \Omega_U \otimes \mathbb{Q})$ is the classical cohomology theory with coefficients in $\Omega_U \otimes \mathbb{Q}$ and $\Omega_U(\mathbb{Z}) \subset \Omega_U \otimes \mathbb{Q}$ is the ring generated by the elements of $\Omega_U \otimes \mathbb{Q}$ with integer Chern numbers. In \cite{ref7} it is shown that $\Omega_U(\mathbb{Z})$ is the ring of polynomials generated by ${[\mathbb{C}P^n] \over n+1}$, $n = 1,2, \dots$ and the Chern-Dold character is defined by the formula $ch_U t = f(u)$, where $f(u)$ is the exponential of the universal formal group law.

\begin{example}
The series
\[ f(u)=\frac{e^{\alpha u}-e^{\beta u}}{\alpha e^{\alpha u}- \beta e^{\beta u}} \]
is the exponential of the formal group
\[ F(t_1,t_2) = \frac{t_1+t_2-a t_1 t_2}{1 - b t_1 t_2},\; \text{ where } a=\alpha+\beta,\; b=\alpha\beta, \]
corresponding to the remarkable two-parameter Todd genus
$T_{\alpha,\beta} \colon \Omega_U \to \mathbb{Z}[a,b]$, a particular case of which are the famous Hirzebruch genera: the Todd genus
($b=0$); the signature ($a=0$); the Eulerian characteristic ($a=2c,\; b=c^2$). See Example (\ref{ex5}) below.
\end{example}

Let $f_k(u) \in A\otimes Q[[u]]$,\, $k=1,2$, be the exponentials of formal group laws $F_1(t_1,t_2)$ and $F_2(t_1,t_2)$ over $A$. The series
$f_1(u)$ and $f_2(u)$ define Hirzebruch genera $L_1$ and $L_2$,
\emph{equivalent} over $A$, if $f_2(u)=\psi(u)f_1(u)$, where
$\psi(u)\in A[[u]]$.

Examples of series $f_1(u)$ and $f_2(u)$ defining equivalent Hirzebruch genera over $A$ such that the formal groups $F_1(t_1,t_2)$ and $F_2(t_1,t_2)$ are not strongly isomorphic can be easily found.
The exponentials $f_1(u)$ and $f_2(u)$ of formal group laws $F_1(t_1,t_2)$ and
$F_2(t_1,t_2)$ define \emph{strongly equivalent} over $A$ Hirzebruch genera $L_1$ and $L_2$ if
$$f_2(u)=\psi(u)f_1(u)\quad \text{ and } \quad f_2(u)=h\big(f_1(u)\big),$$
where $\psi(u)\in A[[u]]$, $h(t) \in A[[t]]$, that is $L_1$ and $L_2$ are equivalent over $A$, and the formal group laws $F_1(t_1,t_2)$ and $F_2(t_1,t_2)$ are strongly isomorphic over $A$.

\begin{example}
The classical $\mathcal{A}$ and $\mathcal{L}$ Hirzebruch genera correspond to the series
\[ f_1(u) = 2\sinh\,\frac{u}{2}\quad {\text and }\quad f_2(u) = \tanh
u, \] which define formal group laws over $\mathbb{Z}[\frac{1}{2}]$:
\begin{align*}
F_1(t_1,t_2) &= t_1 \sqrt{1-\frac{1}{4}t_2^2}+ t_2 \sqrt{1-\frac{1}{4}t_1^2}\,, \\
F_2(t_1,t_2) &= \frac{t_1 + t_2}{1 - t_1 t_2}.
\end{align*}
It is well known that the Hirzebruch genera
$\mathcal{A}$ and $\mathcal{L}$ are strongly equivalent over
$\mathbb{Z}[\frac{1}{2}]$.
\end{example}

\eject

\section{\bf The elliptic formal group law.}

In this section we will solve some problems considered above in the
particular case of a family of formal groups over $E =
\mathbb{Z}[\mu_1, \mu_2, \mu_3, \mu_4, \mu_6]$ defined by the
elliptic curve.
Here $\mu = (\mu_1, \mu_2, \mu_3, \mu_4, \mu_6)$ are the parameters of the elliptic
curve.

\subsection{The elliptic curve.}
Consider the elliptic curve $\mathcal{V}$, given in Weierstrass parametrization by the equation
\begin{equation} \label{Axy}
y^2 + \mu_1 x y + \mu_3 y = x^3 + \mu_2 x^2 + \mu_4 x + \mu_6.
\end{equation}
It is a plane algebraic curve in variables $x,y$, homogeneous in respect to the degrees \\ $\deg x = - 4$, $\deg y = - 6$, $\deg \mu_i = - 2 i$.
Its compactification in $\mathbb{C}P^2$ is given in homogeneous coordinates $(X:Y:Z)$ by the equation
\begin{equation} \label{A1}
Y^2 Z + \mu_1 X Y Z + \mu_3 Y Z^2 = X^3 + \mu_2 X^2 Z + \mu_4 X Z^2 + \mu_6 Z^3.
\end{equation}
The corresponding degrees are $\deg X = - 4$, $\deg Y = - 6$, $\deg Z = 0$, $\deg \mu_i = - 2 i$.

{\bf Note.} There exists also a remarkable parametrisation of an elliptic curve in the Hessian form (see \cite{ref14}):
\[
X^3 + Y^3 + Z^3 = d X Y Z.
\]

Let us fix the Weierstrass form (\ref{A1}) of the equation and now on consider $\mu_i$ as
algebraically independent variables, unless otherwise stipulated.

In $\mathbb{C}P^2$ there exists three canonical coordinate maps.

In the coordinate map $Z \ne 0$ with the coordinates $x = X/Z$ and $y = Y/Z$ the equation of the curve (\ref{A1}) takes the form (\ref{Axy}).
 In this map the curve can be uniformized by the Weierstrass functions (see section 2.6).

In the coordinate map $Y \ne 0$ with the coordinates $t = - X/Y$ and $s = - Z/Y$ the equation of the curve (\ref{A1}) takes the form
\begin{equation} \label{Ats}
s =  t^3 + \mu_1 t s + \mu_2 t^2 s + \mu_3 s^2 + \mu_4 t s^2 + \mu_6 s^3.
\end{equation}
 Here $t$ and $s$ are the
arithmetic Tate coordinates (see \cite{ref16}). Thus $t$ is the uniformizing coordinate for the curve (\ref{Ats}). The remarkable fact about the coordinates $(t, s(t))$ is that the elliptic formal group law in this coordinates is given by series over $E$. We have $\deg t = 2$, $\deg
s = 6$.

In the coordinate map $X \ne 0$ with the coordinates $v = Y/X$ and $w = Z/X$ the equation of the curve (\ref{A1}) takes the form
\begin{equation} \label{Avw}
v w (v + \mu_1 + \mu_3 w) = 1 + \mu_2 w + \mu_4 w^2 + \mu_6 w^3.
\end{equation}
We have $\deg v = -2$, $\deg w = 4$.

The discriminant of the elliptic curve (\ref{Axy}) is $\Delta$, where
\begin{multline} \label{Delta}
4 \Delta = (\mu_1 \mu_3 + 2 \mu_4)^2 (4 \mu_2 + \mu_1^2)^2 - 32 (\mu_1 \mu_3 + 2 \mu_4)^3  - 108 (4 \mu_6 + \mu_3^2)^2 + \\
 + 36 ( \mu_1 \mu_3 + 2 \mu_4) (4 \mu_2 + \mu_1^2) (4 \mu_6 + \mu_3^2) - (4 \mu_2 + \mu_1^2)^3 (4 \mu_6 + \mu_3^2).
\end{multline}
Let
\[
\mu_2 = - (e_3 + e_2 + e_1) - {1 \over 4} \mu_1^2,
\quad
\mu_4 = e_3 e_2 + e_3 e_1 + e_2 e_1 - {1 \over 2} \mu_1 \mu_3,
\quad
\mu_6 = - e_1 e_2 e_3 - {1 \over 4} \mu_3^2.
\]
In this notation the equation (\ref{Axy}) becomes
\[
(2 y + \mu_1 x + \mu_3)^2 = 4 (x - e_1) \, (x - e_2) \, (x - e_3).
\]
We get
\[
\Delta = 16 \, (e_1 - e_2)^2 \, (e_1 - e_3)^2 \, (e_2 - e_3)^2.
\]

\subsection{The geometric group structure on an elliptic curve.} (See \cite{ref17}).

For the geometric group structure on the elliptic curve we have $P_1 + P_2 + P_3 = 0$ for
three points $P_1$, $P_2$, $P_3$ when the points $P_1$, $P_2$ and
$P_3$ lie on a straight line. Let the point $O$ with coordinates
$(0: 1: 0)$ on the elliptic curve be the zero of the group.
Consequently we have $P_1 + P_2 = \overline{P_3}$ in the group when $P_1 + P_2
+ P_3 = 0$ and $P_3 + \overline{P_3} + O = 0$.

The classical geometric group structure on the elliptic curve in Tate coordinates gives a remarkable formal group law $F_{\mu}(t_1, t_2)$ over $E$ that will be called {\it the general elliptic formal group law}. The corresponding formal group law induced by conditions on $\mu_1$, $\mu_2$, $\mu_3$, $\mu_4$, $\mu_6$ will be called {\it the elliptic formal group law}.

In the coordinate map $Y \ne 0$ let $(t_1, s_1)$, $(t_2, s_2)$,
$(t_3, s_3)$ and $(\overline{t_3}, \overline{s_3})$ be the coordinates of the points
$P_1$, $P_2$, $P_3$ and $\overline{P_3}$ respectively. The general elliptic formal group law
$F_{\mu}$ is defined by the condition $F_{\mu}(t_1,t_2) = \overline{t_3}$. In this
coordinates $F_{\mu}(t_1,t_2)$ is a series depending on $t_1$, $t_2$ (see
\cite{ref16}). Due to the construction of the addition law on the elliptic
curve, the series $F_{\mu}(t_1,t_2)$ determines a commutative
one-dimentional formal group. Let $F_{\mu}(t_1, t_2) = t_1 + t_2 +
\sum_{i,j} \alpha_{i,j} t_1^i t_2^j$. It is known that $\alpha_{i,j}
\in E$.

One can find the explicit multiplication law for the Hessian form in \cite{ref14}.

\subsection{The general elliptic formal group law.}

In the coordinate map $Y \ne 0$ the curve $\mathcal{V}$ is given by the equation (\ref{Ats}):
\[
s = t^3 + \mu_1 t s + \mu_2 t^2 s + \mu_3 s^2 + \mu_4 t s^2 + \mu_6 s^3.
\]
The function $s(t)$ is defined by (\ref{Ats}) and the condition $s(0) = 0$. The series expansion at $t=0$ will be
\begin{equation} \label{s}
s =  t^3 + \mu_1 t^4 + (\mu_1^2 + \mu_2) t^5 + (\mu_3 + 2 \mu_2 \mu_1 + \mu_1^3) t^6 + (t^7)
\end{equation}

Let $s = m t + b$ be the equation of the straight line that contains the points $P_1$, $P_2$ and $P_3$ with coordinates $(t_1, s_1)$, $(t_2, s_2)$ and $(t_3, s_3)$ respectively. Using the points $P_1$ and $P_2$ we obtain
\begin{equation} \label{mb}
m = {s_1 - s_2 \over t_1 - t_2}, \qquad b =  {t_1 s_2 - t_2 s_1 \over t_1 - t_2}.
\end{equation}
Using the equation (\ref{Ats}) we get a cubic equation on $t$
\begin{equation} \label{xi}
\xi_0(m) t^3 + \xi_2(m, b) t^2 + \xi_4(m, b) t + \xi_6(b) = 0
\end{equation}
with the roots $t_1$, $t_2$ and $t_3$. Here
\begin{align*}
\xi_0(m) &= 1 + \mu_2 m + \mu_4 m^2 + \mu_6 m^3,
\quad &
\xi_2(m, b) &= \mu_1 m + \mu_2 b + \mu_3 m^2 + 2 \mu_4 m b + 3 \mu_6 m^2 b,
\\
\xi_6(b) &= - b (1 - \mu_3 b - \mu_6 b^2),
\quad &
\xi_4(m, b) &= - m + \mu_1 b + 2 \mu_3 m b + \mu_4 b^2 + 3 \mu_6 m b^2.
\end{align*}

Let
\[
\eta_2(m) = \mu_2 + \mu_4 m + \mu_6 m^2,
\quad
\eta_1(m, b) = \mu_1 + \mu_3 m + \mu_4 b + 2 \mu_6 m b,
\quad
\eta_0(b) = (1 - \mu_3 b - \mu_6 b^2).
\]

Thus
\begin{align*}
\xi_0(m) &= 1 + m \eta_2(m), \quad &
\xi_4(m, b) &= - m \eta_0(b) + b \eta_1(m, b), \\
\xi_2(m, b) &= m \eta_1(m, b) + b \eta_2(m), \quad &
\xi_6(b) &= - b \eta_0(b).
\end{align*}

We get the relations on the coefficients of (\ref{xi})
\begin{equation} \label{help1}
\xi_6(b) - t_1 t_2 \xi_2(m, b) = t_1 t_2 (t_1 + t_2) \xi_0(m),
\end{equation}
\begin{equation} \label{help2}
t_1^2 t_2^2 \xi_0(m) = (t_1 + t_2) \xi_6(b) + t_1 t_2 \xi_4(m, b).
\end{equation}

We have $\xi_0(m) (t_1 + t_2 + t_3) = -  \xi_2(m, b)$, thus
\begin{equation} \label{t3}
t_3 =  - {(t_1 + t_2) \xi_0(m) + \xi_2(m, b) \over \xi_0(m)}.
\end{equation}

Let $s = n t$ be the equation of the straight line that contains the points $P_3$, $\overline{P_3}$ and $O$ with coordinates $(t_3, s_3)$, $(\overline{t_3}, \overline{s_3})$ and $(0, 0)$ respectively. Using the points $P_3$ and $O$ we obtain $n = s_3/t_3 = m + {b \over t_3}$. Using the equation (\ref{Ats}) we get a quadratic equation on $t$
\[
\xi_0(n) t^2 + \xi_2(n, 0) t + \xi_4(n, 0) = 0.
\]
with the roots $t_3$ and $\overline{t_3}$.

Thus \[
\overline{t_3} = - {n \over t_3 \xi_0(n)}.
\]

From (\ref{mb}) we see that $b$, $m$ can be represented as series of $t_1$ and $t_2$ with $\deg b = 6$ and $\deg m = 4$. From (\ref{t3}) we see that $t_3$ can be represented as series of $t_1$ and $t_2$ with $\deg t_3 = 2$. Let us remind $\deg t_i = 2$.

From (\ref{t3}) and (\ref{help1}) have $n = m + t_1 t_2 {(1 + \mu_2 m + \mu_4 m^2 + \mu_6 m^3) \over (1 - \mu_3 b - \mu_6 b^2)}$, thus $n$ can be represented as series of $t_1$ and $t_2$ with $\deg n = 4$.

From this formulas (using (\ref{help2})) we obtain
\[
F_{\mu}(t_1, t_2) =  \left((t_1 + t_2) \eta_0(b) - t_1 t_2 \eta_1(m, b) \right) { \xi_0(m) \over \xi_0(n) \eta_0(b)^2}.
\]
Thus we obtain:

\begin{theorem} The general elliptic formal group law $F_{\mu}(t_1, t_2)$ is given by the formula
\begin{multline} \label{FG}
F_{\mu}(t_1, t_2) = ( t_1 + t_2 - \mu_1 t_1 t_2 - \mu_3 ((t_1+ t_2) b + t_1 t_2 m) - \mu_4 t_1 t_2 b - \mu_6 b ((t_1+ t_2) b + 2 t_1 t_2 m)) \times \\ \times { (1 + \mu_2 m + \mu_4 m^2 + \mu_6 m^3) \over (1 + \mu_2 n + \mu_4 n^2 + \mu_6 n^3) (1 - \mu_3 b - \mu_6 b^2)^2}.
\end{multline}
\end{theorem}

\begin{cor}
Let $F_{\mu}(t_1, t_2) = t_1 + t_2 + \sum_{i,j} \alpha_{i,j} t_1^i t_2^j$ be the general elliptic formal group law. Then $\alpha_{i,j} \in E$.
\end{cor}

Let us denote $p = t_1 t_2$. Using the formulas (\ref{help1}) and (\ref{help2}), we obtain from (\ref{FG})
\begin{equation} \label{first}
F_{\mu}(t_1, t_2) = { (  (t_1 + t_2) \eta_0(b) - p (\mu_1 + \mu_3 m + \mu_4 b + 2 \mu_6 b m) ) \over
\eta_0(b) ( \eta_0(b) + p (\mu_2 + 2 \mu_4 m + 3 \mu_6 m^2) + (\mu_4 +  3 \mu_6 m) p^2 {\xi_0(m) \over \eta_0(b)} + \mu_6 p^3 {\xi_0(m)^2 \over \eta_0(b)^2}))}.
\end{equation}
\begin{example} In the case $(\mu_4, \mu_6) = (0, 0)$ we get
\begin{equation} \label{123}
F_{\mu}(t_1, t_2) =  {  (t_1 + t_2) (1 - \mu_3 b) - \mu_1 t_1 t_2 - \mu_3 t_1 t_2 m \over (1 - \mu_3 b) (1 + \mu_2 t_1 t_2 - \mu_3 b)}.
\end{equation}
\end{example}

Using the formulas (\ref{help1}) and (\ref{help2}), we obtain from (\ref{FG})
\begin{equation} \label{second}
F_{\mu}(t_1, t_2) = {  (t_1 + t_2) (1 + \mu_2 m + \mu_4 m^2 + \mu_6 m^3) + m (\mu_1 + \mu_3 m) + b (\mu_2 + 2 \mu_4 m + 3 \mu_6 m^2) \over  (1 + \mu_2 m + \mu_4 m^2 + \mu_6 m^3) (1 - \mu_3 b) - {b \over p} (\mu_1 + \mu_3 m) (1 - \mu_3 b - \mu_6 b^2) }.
\end{equation}
\begin{example} In the case $(\mu_1, \mu_3) = (0, 0)$ we get
\begin{equation} \label{246}
F_{\mu}(t_1, t_2) = t_1 + t_2 + b {(\mu_2 + 2 \mu_4 m + 3 \mu_6 m^2) \over (1 + \mu_2 m + \mu_4 m^2 + \mu_6 m^3)}.
\end{equation}
\end{example}

We will analyze the general elliptic formal group law (\ref{FG}) and the formal group laws (\ref{123}), (\ref{246}).

\subsection{The equations on the exponentials of the elliptic formal group laws.} \text{ }

Using (\ref{s}) and (\ref{FG}), we get
\begin{equation} \label{dFdt}
 {\partial \over \partial t_2} F_{\mu}(t, t_2) \Big|_{t_2 = 0} = 1 - \mu_1 t - \mu_2 t^2 - 2 \mu_3 s - 2 \mu_4 t s - 3 \mu_6 s^2.
\end{equation}
Let $u = g(t)$ and $\phi(u) = s(f(u))$, where $s(t)$ is defined by (\ref{Ats}) as before. By (\ref{f'}) we come to the system
\begin{equation} \label{fphi}
\left\{ \begin{matrix}
f'(u) =  1 - \mu_1 f(u) - \mu_2 f(u)^2 - 2 \mu_3 \phi(u) - 2 \mu_4 f(u) \phi(u) - 3 \mu_6 \phi(u)^2 \\
\phi(u) = f(u)^3 + \mu_1 f(u) \phi(u) + \mu_2 f(u)^2 \phi(u) + \mu_3 \phi(u)^2 + \mu_4 f(u) \phi(u)^2 + \mu_6 \phi(u)^3. \end{matrix} \right.
\end{equation}
Let $M(t) = 1 - \mu_1 t - \mu_2 t^2$ and $N(t) = \mu_3 + \mu_4 t$. We obtain the following result:

\begin{theorem} Let $(\mu_3, \mu_4, \mu_6) \ne (0, 0, 0)$. Then the exponential $f(u)$ of the elliptic formal group law $F_{\mu}$ is the solution of the equation
\begin{equation} \label{geneq}
\mu_6 \left[ f'^3 + 3 M(f) f'^2 - 4 M(f)^3 + 18 M(f) N(f) f^3 + 27 \mu_6 f^6 \right] = - N(f)^2 \left[ f'^2 - M(f)^2  + 4 N(f) f^3\right],
\end{equation}
with the initial condition $f(0) = 0$ and the condition $f'(0) = 1$ which fixes the branch of solutions.

Let $(\mu_3, \mu_4, \mu_6) = (0, 0, 0)$. Then the exponential $f(u)$ of the elliptic formal group law $F_{\mu}$ is the solution of the equation
\begin{equation} \label{Ricatti}
f' = M(f)
\end{equation}
with the initial condition $f(0) = 0$.
\end{theorem}

\begin{example} Let $\mu_6 = 0$, $(\mu_3, \mu_4) \ne (0, 0)$. Then the equation (\ref{geneq}) gives
\begin{equation} \label{Erm}
f'^2 = 1 - 2 \mu_1 f + (\mu_1^2 - 2 \mu_2) f^2 + (2 \mu_1 \mu_2 - 4 \mu_3) f^3 + (\mu_2^2 - 4 \mu_4) f^4.
\end{equation}
\end{example}
Remark that for $\mu_3 = \mu_4 = 0$ this equation leads to the equation (\ref{Ricatti}) (the sign is determined by the conditions $f(0) = 0$, $f'(0) = 1$), so the equation (\ref{Erm}) should be considered as the general equation on the exponential in the case $\mu_6 = 0$.

\begin{example}
Let $(\mu_1, \mu_2, \mu_3) = (0, 0, 0)$. Then the equation (\ref{geneq}) gives
\begin{equation} \label{46}
\mu_6 f'^3 + (\mu_4^2 f^2 + 3 \mu_6) f'^2 + (4 \mu_4^3 + 27 \mu_6^2) f^6 + 18 \mu_4 \mu_6 f^4 - \mu_4^2 f^2 - 4 \mu_6 = 0.
\end{equation}
\end{example}

\subsection{The elliptic curve in the standard Weierstrass form.} \text{ }

An elliptic curve in the {\it standard Weierstrass form} is given by the equation
\begin{equation} \label{Wei}
y^2 = 4 x^3 - g_2 x - g_3.
\end{equation}
There is the classical Weierstrass function $\sigma(u) = \sigma(u; g_2, g_3)$ related with this curve (see \cite{ref15}). It is an entire odd function of $u \in \mathbb{C}$ such that $\sigma(u) = u+(u^5)$. It is a quasiperiodic function with the periods $2 \omega_1$, $2 \omega_2$:
\[
\sigma(u+2\omega_k) = -\exp\big( 2\eta_k(u+\omega_k)
\big)\sigma(u), \quad k=1,2.
\]

The Weierstrass functions $\zeta(u) = \zeta(u; g_2, g_3)$ and $\wp(u) = \wp(u; g_2, g_3)$ are defined by the equations
\begin{equation}\label{zeta}
\zeta(u) =(\ln \sigma(u))' \quad \text{ and } \quad \wp(u) = -\zeta(u)'.
\end{equation}
We have $\lim_{u \to 0} (\wp(u) - {1 \over u^2}) = 0$.
The map $u \mapsto (x,y) = (\wp(u), \wp'(u))$ gives the {\it Weierstrass uniformization}
\begin{equation}\label{W}
\wp(u)'^2 = 4 \wp(u)^3 - g_2 \wp(u) - g_3
\end{equation}
 of the elliptic curve in the standard Weierstrass form.

In the case $\mu_1 = \mu_2 = \mu_3 = 0$ the elliptic curve in Tate coordinates (\ref{Ats}) is given by the equation
\begin{equation} \label{tsg}
s = t^3 + \mu_4 t s^2 + \mu_6 s^3.
\end{equation}
It follows from (\ref{246}) that the formal group law takes the form
\begin{equation} \label{F1}
F_{\mu}(t_1, t_2) = t_1 + t_2 + b m {2 \mu_4 + 3 \mu_6 m \over 1 + \mu_4 m^2 + \mu_6 m^3}
\end{equation}
and therefore
\[
{\partial F_{\mu}(t, t_2) \over \partial t_2} |_{t_2 = 0} = \rho(t), \quad \text{where} \quad \rho(t) = 1 - 2 \mu_4 t s(t) - 3 \mu_6 s(t)^2.
\]
Using (\ref{f'}), we get
\begin{equation} \label{f1}
f'(u) = \rho(f(u)).
\end{equation}

In the equation (\ref{Axy}) for $x = \widetilde{x}$, $2 y = \widetilde{y}$, $(\mu_1, \mu_2, \mu_3) = (0,0,0)$, $- 4 \mu_4 = g_2$, $- 4 \mu_6 = g_3$, we obtain the standard Weierstrass equation
\[
\widetilde{y}^2 = 4 \widetilde{x}^3 - g_2 \widetilde{x} - g_3.
\]
It is equivalent to the equation (\ref{tsg}) with $t = - 2 \widetilde{x}/\widetilde{y}$ and $s = - 2/\widetilde{y}$.

Using the Weierstrass uniformization (\ref{W}) of (\ref{Wei}), we get the uniformization \\
$u \mapsto (t =  {- 2 \wp(u) \over \wp'(u)}, s = {- 2 \over \wp'(u)})$ of (\ref{tsg}).

\begin{lemma}
The function
\begin{equation} \label{fg}
f(u) = {- 2 \wp(u; g_2, g_3) \over \wp'(u; g_2, g_3)}
\end{equation}
is the exponential of the formal group (\ref{F1}), where $g_2 = - 4 \mu_4$ and $g_3 = - 4 \mu_6$.
\end{lemma}

Thus the Weierstrass uniformization induces a strong isomorphism of the linear group with the coordinate $u$ and the formal group (\ref{F1}), corresponding to the elliptic curve with Tate coordinates for $\mu_1=\mu_2=\mu_3 = 0$.

{\bf Proof.} We have $t(u) = {- 2 \wp(u; g_2, g_3) \over \wp'(u; g_2, g_3)}$, $t(0) = 0$, $t'(0) = 1$. Using $s(u) = {- 2 \over \wp'(u)}$ and (\ref{W}), we see that $t(u)$ gives a solution of the equation (\ref{f1}), thus $t(u)$ is the exponential of the formal group (\ref{F1}).

\begin{cor}
We have
\[
{- 2 \wp(u+v; g_2, g_3) \over \wp'(u+v; g_2, g_3)} = F_g({- 2 \wp(u; g_2, g_3) \over \wp'(u; g_2, g_3)}, {- 2 \wp(v; g_2, g_3) \over \wp'(v; g_2, g_3)}),
\]
where
\begin{equation} \label{Fg}
F_{g}(t_1, t_2) = t_1 + t_2 - b m {2 g_2 + 3 g_3 m \over 4 - g_2 m^2 - g_3 m^3}.
\end{equation}
\end{cor}

\subsection{The reduction to the standard Weierstrass curve.} \text{}

Consider the following linear transformation of $\mathbb{C}P^2$
\[
(X:Y:Z) \mapsto (\widetilde{X}: \widetilde{Y} : \widetilde{Z}) = (X + {1 \over 12} (4 \mu_2 + \mu_1^2) Z: 2 Y + \mu_1 X + \mu_3 Z : Z).
\]

It brings the curve $\mathcal{V}$ to the curve $\widetilde{\mathcal{V}}$ given by the equation
\[
\widetilde{Y}^2 \widetilde{Z} = 4 \widetilde{X}^3 - g_2 \widetilde{X} \widetilde{Z}^2 - g_3 \widetilde{Z}^3,
\]
where
\begin{align} \label{g2}
g_2 & = {1 \over 12}(4 \mu_2 + \mu_1^2)^2 - 2 (\mu_1 \mu_3 + 2 \mu_4),
\\ \label{g3}
g_3 & = {1 \over 6} ( \mu_1 \mu_3 + 2 \mu_4) (4 \mu_2 + \mu_1^2) - {1 \over 6^3} (4 \mu_2 + \mu_1^2)^3 - 4 \mu_6 - \mu_3^2.
\end{align}
Notice that this linear transformation brings $O$ to $O$, so it gives a homomorphism of formal groups $F_{\mu} \mapsto F_g$, where $F_g$ is the formal group (\ref{Fg}) given by the geometric structure of the group on the elliptic curve $\widetilde{\mathcal{V}}$.

In the coordinate map $Y \ne 0$ we have the Tate coordinates $\widetilde{t} = - 2 \widetilde{X}/\widetilde{Y}$, $\widetilde{s} = - 2 \widetilde{Z}/\widetilde{Y}$, and the curve $\widetilde{\mathcal{V}}$ is given by the equation
\[
\widetilde{s} = \widetilde{t}^3 - {1 \over 4} g_2 \widetilde{t} \widetilde{s} - {1 \over 4} g_3 \widetilde{s}^3.
\]
We come to equation (\ref{tsg}) for $g_2 = - 4 \mu_4$ and $g_3 = - 4 \mu_6$.

\begin{lemma} The Tate coordinates $t$ and $s$ of the curve $\mathcal{V}$ are connected to the Tate coordinates $\widetilde{t}$ and $\widetilde{s}$ of the curve $\widetilde{\mathcal{V}}$ by the following formulas:
\[
t = {\widetilde{t} - {1 \over 12}(4 \mu_2 + \mu_1^2) \widetilde{s} \over 1 + {1 \over 2} \mu_1 \widetilde{t} - {1 \over 2} ({1 \over 12} \mu_1 (4 \mu_2 + \mu_1^2) - \mu_3) \widetilde{s} },
\quad s = {\widetilde{s} \over 1 + {1 \over 2} \mu_1 \widetilde{t} - {1 \over 2} ({1 \over 12} \mu_1 (4 \mu_2 + \mu_1^2) - \mu_3) \widetilde{s}}.
\]
\end{lemma}

Let
\[
\psi(t) = \widetilde{t}(t) = {t + {1 \over 12} (4 \mu_2 + \mu_1^2) s(t) \over 1 - {1 \over 2} \mu_1 t - {1 \over 2} \mu_3 s(t)}.
\]
Thus
\begin{equation} \label{psi}
\psi(F_{\mu}(t_1, t_2)) = F_g(\psi(t_1), \psi(t_2)).
\end{equation}

In the coordinate map $Z \ne 0$ we have
\begin{equation} \label{FORM}
\widetilde{x} = x + {1 \over 12}(4 \mu_2 + \mu_1^2), \qquad \widetilde{y} = 2 y + \mu_1 x + \mu_3.
\end{equation}
The curve $\widetilde{\mathcal{V}}$ is given by the equation
\[
\widetilde{y}^2 = 4 \widetilde{x}^3 - g_2 \widetilde{x} - g_3.
\]
Using the Weierstrass uniformization of the curve $\widetilde{\mathcal{V}}$: $(\widetilde{x}, \widetilde{y}) =
(\wp(u; g_2, g_3), \wp'(u, g_2, g_3))$, we obtain the uniformization of the curve $\mathcal{V}$:
\[
x = \wp(u; g_2, g_3) -  {1 \over 12}(4 \mu_2 + \mu_1^2), \quad y = {1 \over 2} (\wp'(u, g_2, g_3) - \mu_1 \wp(u; g_2, g_3) + {1 \over 12} \mu_1 (4 \mu_2 + \mu_1^2) - \mu_3).
\]

Using (\ref{fg}) and (\ref{psi}), we get:

\begin{cor}
 The exponential of the general elliptic formal group (\ref{FG}) is
\begin{equation} \label{exp}
f(u) = - 2 { \wp(u; g_2, g_3) -  {1 \over 12}(4 \mu_2 + \mu_1^2) \over \wp'(u; g_2, g_3) - \mu_1 \wp(u; g_2, g_3) + {1 \over 12} \mu_1 (4 \mu_2 + \mu_1^2) - \mu_3},
\end{equation}
where $g_2$ and $g_3$ are given by (\ref{g2}) and (\ref{g3}).
\end{cor}

Thus the Weierstrass uniformization of the curve (\ref{Wei}) induces a strong isomorphism of the linear group with the coordinate $u$ and the formal group (\ref{FG}), corresponding to the elliptic curve with Tate coordinates (\ref{Ats}).

\begin{cor}
The solution $f(u)$ of the equation (\ref{geneq}) with conditions $f(0) = 0$, $f'(0) = 1$ is given by formula (\ref{exp}).
\end{cor}

\begin{cor}
 $f(u) \in H E[[u]]$.
\end{cor}

\begin{cor}
We have
\begin{equation} \label{exp2}
{1 \over f(u)} = - {1 \over 2} {\wp'(u) + \wp'(w) \over \wp(u) - \wp(v)} + {\mu_1 \over 2},
\end{equation}
where $\wp(u) = \wp(u; g_2, g_3)$, $\wp'(w) = - \mu_3$ and $\wp(v) = {1 \over 12}(4 \mu_2 + \mu_1^2)$.
\end{cor}

\begin{cor}
The exponential of the elliptic formal group law in the non-degenerate case is the ellipitc function of order $2$ iff $\mu_6 = 0$. It is the ellipitc function of order $3$ in the general non-degenerate case.
\end{cor}

We will consider the elliptic sine $f(u) = sn(u; \delta, \varepsilon)$ as the solution of the equation
\[
f(u)'^2 = R(f(u)), \quad R(t) = 1 - 2 \delta t^2 + \varepsilon t^4
\]
with conditions $f(0) = 0$, $f'(0) = 1$. It has the classical addition law
\begin{equation} \label{cl}
F(t_1, t_2) = {t_1 \sqrt{R(t_2)} + t_2 \sqrt{R(t_1)} \over 1 - \varepsilon t_1^2 t_2^2}
\end{equation}
\begin{example}
The exponential of the elliptic formal group law is the elliptic sine with parametra $(\delta, \varepsilon)$ if and only if $(\mu_1, \mu_3, \mu_6) = (0, 0, 0)$, $\mu_2 = \delta$, $\mu_4 = {1 \over 4} (\delta^2 - \varepsilon)$.
Thus the classical addition law (\ref{cl}) gives a formal group over $\mathbb{Z}[\mu_2, \mu_4]$.
\end{example}

\subsection{The 2-height of the elliptic formal group laws.} \text{ }

For any formal group $F(t_1, t_2)$ over $A$ the formula
\[
F(t,t) = t^{2^h} + ... (\text{mod} \, 2)
\]
holds for some $h \geq 1$. Such number $h$ is called the {\it 2-height} of the  formal group $F$.

 Let us find $h$ for the elliptic formal group laws.

Over the ring $\mathbb{Z}_2[\mu_i][[t]]$ for $t_1 = t_2 = t$ we have
\[
m = s'(t) = { t^2 + \mu_1 s(t) + \mu_4 s(t^2) \over 1 + \mu_1 t + \mu_2 t^2 + \mu_6 s(t^2)},
\quad
b = t s'(t) + s(t) = { \mu_1 t s(t) + \mu_3 s(t^2) \over 1 + \mu_1 t + \mu_2 t^2 + \mu_6 s(t^2)},
\]
\[
n = s'(t) + t^2 {(1 + \mu_2 m + \mu_4 m^2 + \mu_6 m^3) \over (1 + \mu_3 b + \mu_6 b^2)}.
\]

\begin{lemma}
 Over the ring $\mathbb{Z}_2[\mu_i]$ we have
\begin{equation} \label{form}
F_{\mu}(t, t) = (  \mu_1 t^2 + \mu_3 t^2 m + \mu_4 t^2 b) { (1 + \mu_2 m + \mu_4 m^2 + \mu_6 m^3) \over (1 + \mu_2 n + \mu_4 n^2 + \mu_6 n^3) (1 - \mu_3 b - \mu_6 b^2)^2}.
\end{equation}
\end{lemma}

Thus, we have:

\begin{cor}
For $\mu_1 \ne 0$ the height is $1$.

For $\mu_1 = 0$, the formula (\ref{form}) takes the form
\[
F_{\mu}(t, t) = {\mu_3 t^4 \over 1 - \mu_2 t^2 - \mu_6 s(t^2)} { (1 + \mu_2 m + \mu_4 m^2 + \mu_6 m^3) \over (1 + \mu_2 n + \mu_4 n^2 + \mu_6 n^3) (1 - \mu_3 b - \mu_6 b^2)^2}.
\]
Thus in the case $\mu_1 = 0$, $\mu_3 \ne 0$ the height is $2$ and the elliptic curve is supersingular.

For $\mu_1 = 0$, $\mu_3 = 0$ we see the height is $\infty$.
\end{cor}

\subsection{The formal group law over the ring with trivial multiplication.} \text{ }

We will describe the formal group law modulo the ideal of decomposable element in the ring $E$
in order to get important information on the homomorphism $\phi: \mathcal{A} \to E$.

Let $E^{(1)} = E / (\widetilde{E})^2$, where $\widetilde{E} = Ker(E \mapsto \mathbb{Z}: \mu_i  \mapsto 0)$.
Over the ring $E^{(1)}$ from (\ref{Ats}) we have
\[
s =  t^3 + \mu_1 t^4 + \mu_2 t^5 + \mu_3 t^6 + \mu_4 t^7 + \mu_6 t^9.
\]
From (\ref{FG}) we get
\begin{multline*}
F_{\mu}(t_1, t_2) =  t_1 + t_2 - t_1 t_2 \left[ \mu_1 + \mu_2 (t_1 + t_2) + \mu_3 (2 t_1^2 + 3 t_1 t_2 + 2 t_2^2) + \right. \\ \left. + 2 \mu_4 (t_1 + t_2) (t_1^2 + t_1 t_2 + t_2^2) + 3 \mu_6  (t_1 + t_2) (t_1^2 + t_1 t_2 + t_2^2)^2 \right].
\end{multline*}
It follows from (\ref{f'}) that
\[
f'(t) = {\partial F_{\mu}(t_1, t_2) \over \partial t_2} |_{t_2 = 0} = 1 - (\mu_1 t_1 + \mu_2 t_1^2 + 2 \mu_3 t_1^3 + 2 \mu_4 t_1^4 + 3 \mu_6 t_1^6).
\]
Thus, over $E^{(1)}$ the formula holds:
\[
f(t) = t - {1 \over 2} \mu_1 t^2 - {1 \over 3} \mu_2 t^3 - {1 \over 2} \mu_3 t^4 - {2 \over 5} \mu_4 t^5 - {3 \over 7} \mu_6 t^7.
\]
We get $f_1 = - {1 \over 2} \mu_1$, $f_2 = - {1 \over 3} \mu_2$, $f_3 = - {1 \over 2} \mu_3$, $f_4 = - {2 \over 5} \mu_4$, $f_6 = - {3 \over 7} \mu_6$.
Notice that $\nu(2) = 2$, $\nu(3) = 3$, $\nu(4) = 2$, $\nu(5) = 5$, $\nu(7) = 7$.

Using the description of the multiplicative generators of the ring $\mathcal{A}$ (see page 4), we obtain
\[
a_n^* = \nu(n+1) b_n^*
\quad
\text{and}
\quad
\phi(b_n) = f_n.
\]
Let $E_F$ be the subring of $E$, generated by the coefficients of $F$. Therefore\\
1. The composition of the maps $E_F \hookrightarrow E$ and $E \to \mathbb{Z}[\mu_1, \mu_2, \mu_3]$: $\mu_i \mapsto \mu_i$, \\ $i = 1,2,3$, $\mu_i \mapsto 0$, $i = 4,6$ is an epimorphism.\\
2. The composition of the maps $(E_F)_{(p)} \hookrightarrow E_{(p)}$ and $E_{(p)} \to \mathbb{Z}_{(p)}[\mu_1, \mu_2, \mu_3, \mu_6]$: $\mu_i \mapsto \mu_i$, $i = 1,2,3,6$, $\mu_4 \mapsto 0$, is an epimorphism if $p \ne 3$.\\
3. The composition of the maps $(E_F)_{(p)} \hookrightarrow E_{(p)}$ and $E_{(p)} \to \mathbb{Z}_{(p)}[\mu_1, \mu_2, \mu_3, \mu_4]$: $\mu_i \mapsto \mu_i$, $i = 1,2,3,4$, $\mu_6 \mapsto 0$, is an epimorphism if $p \ne 2$.\\
In particular, the coefficients of the formal group law $F(t_1, t_2)$ generate multiplicatively the ring $E_{(p)}$ for $p \ne 2, 3$.

\subsection{Authomorphisms of elliptic formal group laws.} \text{ }

It is wellknown that the group structure on the elliptic curve can have non-trivial automorphisms of order only $2, 3, 4,$ and $6$. We will derive this result using the formal group law in Tate coordinates. We included the exposition of this result because it will be used below.

Let us consider the formal group law $F_{\mu}$ over $E \otimes \left( \mathbb{Z}[\alpha] / J \right)$, where $J$ is some ideal.

Any linear authomorphism of a formal group law is given by the identity
\begin{equation} \label{Aut}
F_{\mu}(\alpha t_1, \alpha t_2) = \alpha F_{\mu}(t_1, t_2).
\end{equation}

Using (\ref{f'}) we obtain
\[
g'(\alpha t) = g'(t).
\]
Thus $g(\alpha t) = \alpha g(t)$ and $f(\alpha t) = \alpha f(t)$.

Because $g(t) = t + ...$ is a series of $t$, it can be presented in the form $g(t) = t \psi(t^n)$ for some $n$ and some series $\psi(t)$.

In the case $\psi(t^n) = 1$ we get $g(t) = t$, so $F_{\mu}(t_1,t_2) = t_1 + t_2$ is a linear group and the identity (\ref{Aut}) is valid for any ideal $J$.

For $\psi \ne 1$ fix the maximal $n$ in the form $g(t) = t \psi(t^n)$. Then it follows from the identity $g(\alpha t) = \alpha g(t)$ that $(\alpha^n - 1) \in J$. Thus ${1 \over g'(t)}$ can be presented in the form $\phi(t^n)$ for some series $\phi(t)$.

Let us consider the case of elliptic formal group laws.

Using (\ref{dFdt}) and (\ref{f'}), we get the condition $\rho(\alpha t) = \rho(t)$ for
\begin{equation} \label{1/g'}
\rho(t) = 1 - \mu_1 t - \mu_2 t^2 - 2 \mu_3 s - 2 \mu_4 t s - 3 \mu_6 s^2.
\end{equation}

Let $n=2$. The function $\rho(t) = \phi(t^2)$ should be an even function of $t$, thus $\mu_1 = \mu_3 = 0$.
In this case
\[
\rho(t) = 1 - \mu_2 t^2 - 2 \mu_4 t s - 3 \mu_6 s^2,
\]
where $s(t)$ determined by the relation $s = t^3 + \mu_2 t^2 s + \mu_4 t s^2 + \mu_6 s^3$ is an odd function. See Example \ref{ex1}.

Let $n=3$. Then $\rho(t) = \phi(t^3)$, thus $\mu_1 = \mu_2 = \mu_4 = 0$.
In this case
\[
\rho(t) = 1 - 2 \mu_3 s - 3 \mu_6 s^2,
\]
where $s(t)$ determined by the relation $s =  t^3 + \mu_3 s^2 + \mu_6 s^3$, thus $s$ is a function of $t^3$. See Example \ref{ex2}.

Let $n=4$. Then $\rho(t) = \phi(t^4)$, thus $\mu_1 = \mu_2 = \mu_3 = \mu_6 = 0$.
In this case
\[
\rho(t) = 1 - 2 \mu_4 t s,
\]
where $t s(t)$ determined by the relation $t s =  t^4 + \mu_4 t^2 s^2$, thus $t s$ is a function of $t^4$. See Example \ref{ex3}.

Let $n=6$. Then ${1 \over g'(t)} = \phi(t^6)$, thus $\mu_1 = \mu_2 = \mu_3 = \mu_4 = 0$.
In this case
\[
\rho(t) = 1 - 3 \mu_6 s^2,
\]
where $s(t)$ determined by the relation $s =  t^3 + \mu_6 s^3$, thus $s(t)$ is an odd function of $t^3$ and $s^2(t)$ is a function of $t^6$. See Example \ref{ex4}.

Let $n=5$ or $n\geq7$. Then we should have $\rho(t) = \phi(t^n)$, but it follows from (\ref{1/g'}) that $\mu_1 = \mu_2 = \mu_3 = \mu_4 = \mu_6 = 0$, thus $g'(t) = 1$ and $\psi(t^n) = 1$.

\subsection{Differential equations, connecting the Tate coordinates \\ of the elliptic curve with its parameters.} \text{ }

Using the relation (\ref{Ats})
\[
s = t^3 + \mu_1 t s + \mu_2 t^2 s + \mu_3 s^2 + \mu_4 t s^2 + \mu_6 s^3,
\]
one can consider $s$ as a function $s(t;\mu)$ of $t$ and $\mu = (\mu_1, \mu_2, \mu_3, \mu_4, \mu_6)$. Then
\begin{equation} \label{dsdt}
(1 - \mu_1 t - \mu_2 t^2 - 2 \mu_3 s - 2 \mu_4 t s - 3 \mu_6 s^2) {\partial s \over \partial t} = 3 t^2 + \mu_1 s + 2 \mu_2 t s + \mu_4 s^2.
\end{equation}

\begin{lemma}
Let $\mu(v) = (0, 3 v + c_2, c_3, 3 v^2 + 2 c_2 v + c_4, v^3 + c_2 v^2 + c_4 v + c_6)$, where $c_k$ do not depend on $v$, $k = 2,3,4,6$. The function $S(t,v) = s(t, \mu(v))$ satisfies the Hopf equation
\begin{equation} \label{Hopf}
{\partial S \over \partial v} = S {\partial S \over \partial t}
\end{equation}
with the initial conditions $S(t,0) = s_0(t)$, where $s_0(t)$ is defined by the equation
\[
s_0 = t^3 + c_2 t^2 s_0 + c_3 s_0^2 + c_4 t s_0^2 + c_6 s_0^3.
\]
\end{lemma}

{\bf Proof.} Consider the path $\mu: \mathbb{C} \to \mathbb{C}^5$: $v \mapsto \mu(v)$, where $\mu_i(v)$ are given in the lemma. Thus
\[
{\partial s \over \partial v} (1 - \mu_1 t - \mu_2 t^2 - 2 \mu_3 s - 2 \mu_4 t s - 3 \mu_6 s^2) = 3 t^2 s + 2 (3 v + c_2) t s^2 + (3 v^2 + 2 c_2 v + c_4) s^3.
\]
Comparing with (\ref{dsdt}), we come to the Hopf equation (\ref{Hopf}).

{\bf Note:}
Let $U=U(\tau,\upsilon;\alpha)$ be the solution of the Hopf equation
\[ \frac{\partial U}{\partial\upsilon}=U\frac{\partial}{\partial\tau}U \]
with $U(\tau,0;\alpha)=\frac{\tau^2}{1-\alpha\tau}$.

Then $U(\tau,\upsilon;\alpha)=\sum\limits_{n\geqslant0}f(As^n)\tau^{n+2}$
where $f(As^n)=\alpha^n+f_{n-1}\alpha^{n-1}\upsilon+\ldots+f_0\upsilon^n$.
Here $As^n$ is $n$-dimensional associahedron, Stasheff polytope
$K_{n+2}$, and $f_k=f_k(As^n)$ is the number of $k$-dimensional
faces. Then the function $U$ satisfies the equation (see \cite{ref8})
\[ \upsilon(\alpha+\upsilon)U^2-\big(1- (\alpha+2\upsilon)\tau \big)U+\tau^2=0. \]

\begin{lemma}
The path $\mu(v) $, where $ \mu_1(v) = 0$, $\mu_2(v) = 3 v + c_2$, \\ $\mu_3(v) = c_3$, $\mu_4(v) = 3 v^2 + 2 c_2 v + c_4$, $\mu_6(v) = v^3 + c_2 v^2 + c_4 v + c_6$, defines a family of elliptic curves with the same standard Weirstrass form for any $v$.
\end{lemma}

{\bf Proof.} The reduction to the standard Weierstrass form gives a mapping \\
$
\mathbb{C}^5 \to \mathbb{C}^2:$ $ \mu \mapsto (g_2, g_3),
$
 defined by (\ref{g2}), (\ref{g3}). \\
Direct calculations give $g_2(\mu(v)) = g_2(\mu(0))$ and $g_3(\mu(v)) = g_3(\mu(0))$.

\begin{lemma}
The function $S(\tau,v) = s(t, \mu(v))$, where $\tau = t^3$, $ \mu_1(v) = \mu_2(v) = \mu_4(v) = 0$, $\mu_3(v) = \alpha v + c_3$, $\mu_6(v) = \beta v + c_6$, satisfies the equation
\begin{equation} \label{Hopf2}
{\partial S \over \partial v} = (\alpha S^2 + \beta S^3) {\partial S \over \partial \tau}
\end{equation}
with the initial conditions $S(\tau,0) = s_0(\tau)$, where $s_0(\tau)$ is defined by the equation
\[
s_0 = \tau + c_3 s_0^2 + c_6 s_0^3.
\]
\end{lemma}

{\bf Proof.} Consider the path $\mu: \mathbb{C} \to \mathbb{C}^5$: $v \mapsto \mu(v)$, where $\mu_i(v)$ are given in the lemma.
We have
\[
s = t^3 + \mu_3 s^2 + \mu_6 s^3,
\]
so $s(t, \mu(v))$ depends on $\tau = t^3$ .
Thus for $S(\tau, v)$ we have
\[
{\partial S \over \partial \tau} (1 - 2 \mu_3 S - 3 \mu_6 S^2) = 1,
\]
\[
{\partial S \over \partial v} (1 - 2 \mu_3 S - 3 \mu_6 S^2) = \alpha S^2 + \beta S^3,
\]
and we come to the equation (\ref{Hopf2}).

\section{\bf Hurwitz series defined by elliptic curves.}

\subsection{The sigma-function of the elliptic curve.}

The sigma function $\sigma(u)$ has a series expansion in powers of $u$ over the polynomial ring $\mathbb{Q}[g_2, g_3]$ in the vicinity of $u=0$. An initial segment of the series has the form
\begin{equation}\label{F-2}
\sigma(u)=  u - {g_2 \, u^5 \over 2 \cdot 5!} - {6 \, g_3 \, u^7 \over 7!} -
{ g_2^2 \, u^9 \over 4 \cdot 8!} - { 18 \, g_2 \, g_3 \, u^{11} \over {11}!} +
(u^{13}).
\end{equation}

\begin{theorem}
The sigma function $\sigma(u)$ is a Hurwitz series over $\mathbb{Z}[{1 \over 2}][g_2, g_3]$.
\end{theorem}

For the proof see \cite{ref13}.

The following operators annihilate the sigma function
\[
Q_0 = 4 g_2 {\partial \over \partial g_2} + 6 g_3 {\partial \over \partial g_3} - u {\partial \over \partial u} + 1,
\quad
Q_2 = 6 g_3  {\partial \over \partial g_2} + {1 \over 3} g_2^2 {\partial \over \partial g_3} - {1 \over 2} {\partial^2 \over
\partial u^2} - {1 \over 24} g_2 u^2.
\]

\begin{theorem}
Set
\[
\sigma(u) = u \sum_{i, j \geq 0} {a_{i,j} \over (4 i + 6 j + 1)!} ({g_2 u^4 \over 2})^i (2 g_3 u^6)^j.
\]
The Weierstrass recursion for the sigma function is given by the formulas
\[
a_{i,j} = 3 (i+1) a_{i+1,j-1} + {16 \over 3} (j+1) a_{i-2,j+1} - {1 \over 3} (4 i + 6 j -1) (2 i + 3 j - 1) a_{i-1,j} \text{ for }  i \geq 0, \; j  \geq 0, \; (i,j) \neq (0,0),
\]
\[
a_{0,0} = 1, \quad a_{i,j} = 0 \text{ for }  i < 0 \text{ or } j < 0,
\]
which define $a_{i,j}$ for $2 i + 3 j \leq 0$, and if $a_{i,j}$ is
defined for $2 i + 3 j < m$ where $m>0$, then $a_{i,j}$ is defined
recursively for $2 i + 3 j = m$.
\end{theorem}

{\bf Proof.}
Consider the equation $Q_2 \sigma(u) = 0$. It follows that
\[
 18 (i+1) \sum_{i \geq 0, j \geq 1} {a_{i+1,j-1} \over (4 i + 6 j -1)!} ({g_2 u^4 \over 2})^{i} (2 g_3 u^6)^{j} + 32 (j+1) \sum_{i \geq 2, j \geq 0} {a_{i-2,j+1} \over (4 i + 6 j - 1)!} ({g_2 u^4 \over 2})^{i} (2 g_3 u^6)^{j} - \] \[ - 6 (4 i + 6 j + 1) (4 i + 6 j) \sum_{i \geq 0, j \geq 0} {a_{i,j} \over (4 i + 6 j+1)!} ({g_2 u^4 \over 2})^i (2 g_3 u^6)^j - \sum_{i \geq 1, j \geq 0} {a_{i-1,j} \over (4 i + 6 j - 3)!} ({g_2 u^4 \over 2})^{i} (2 g_3 u^6)^j = 0.
\]

For $i = j = 0$ we get $0 = 0$,
\begin{align*}
\text{for $j = 0, i = 1$:} & \qquad  a_{1,0} = - a_{0,0},
\\
\text{for $i = 0, j \geq 1$:} & \qquad a_{0,j} = 3 a_{1,j-1},
\\
\text{for $j = 0, i \geq 2$:} & \qquad 3 a_{i,0} = 16 a_{i-2, 1} - (2 i  - 1) (4 i  - 1) a_{i-1,0},
\\
\text{for $i = 1, j \geq 1$:} & \qquad a_{1,j} = 6 a_{2,j-1} - ( 2 j + 1) ( 3 j + 1) a_{0,j},
\\
\text{for $i \geq 2, j \geq 1$:} & \qquad 3 a_{i,j} = 9 (i+1) a_{i+1,j-1} + 16 (j+1) a_{i-2,j+1} - (4 i + 6 j -1) (2 i + 3 j - 1) a_{i-1,j}.
\end{align*}

Put $a_{i,j} = 0$ for $i < 0$ or $j < 0$. For all $(i,j) \neq (0,0)$ the formula holds:
\[
a_{i,j} = 3 (i+1) a_{i+1,j-1} + {16 \over 3} (j+1) a_{i-2,j+1} - {1 \over 3} (4 i + 6 j -1) (2 i + 3 j - 1) a_{i-1,j}.
\]
The definition of the sigma function gives the initial condition for the recursion
$a_{0,0} = 1$.

Thus $a_{i,j} \in \mathbb{Z}[{1 \over 3}]$ and we obtain

\begin{cor} The sigma function is a Hurwitz series over $\mathbb{Z}[{1 \over 3}, {g_2 \over 2}, 2 g_3]$:
\[
\sigma(u) \in H \mathbb{Z}[{1 \over 3}, {g_2 \over 2}, 2 g_3] [[u]].
\]
\end{cor}

From $\sigma(u) \in H \mathbb{Z}[{1 \over 3}, {g_2 \over 2}, 2 g_3] [[u]]$ and $\sigma(u) \in H \mathbb{Z}[{1 \over 2}, g_2 , g_3] [[u]]$ we obtain:

\begin{theorem} The sigma function is a Hurwitz series over $\mathbb{Z}[{g_2 \over 2}, 2 g_3]$:
\[
\sigma(u) \in H \mathbb{Z}[{g_2 \over 2}, 2 g_3] [[u]],
\] that is $a_{i,j} \in \mathbb{Z}$.
\end{theorem}

Explicitly, we have
\[
a_{0,0} = 1, \; a_{1,0} = -1, \; a_{2,0} = - 3^2, \; a_{3,0} = 3 \cdot 23, \; a_{4,0} = 3 \cdot 107, \;
\]
\[
a_{0,1} = -3, \; a_{1,1} = -2 \cdot 3^2, \; a_{2,1} = 3^3 \cdot 19, \; a_{3,1} = 2^2 \cdot 3^3 \cdot 311, \; a_{4,1} = 3^3 \cdot 5 \cdot 20807,
\]
\[
a_{0,2} = -2 \cdot 3^3, \; a_{1,2} = 2^3 \cdot 3^3 \cdot 23, \; a_{2,2} = 2^2 \cdot 3^5 \cdot 5 \cdot 53, \; a_{3,2} = 2^3 \cdot 3^4 \cdot 5 \cdot 37 \cdot 167, \; a_{4,2} = - 2 \cdot 3^6 \cdot 5 \cdot 17 \cdot 3037.
\]

Let $b_{i,j} = 2^{3 i + 4 j} \, 3^{i+j} \, {i! \, j! \over (4i + 6j + 1)!} \, a_{i,j}$; $b_{i,j} \in \mathbb{Q}$.
We get

\[
(4i + 6j + 1) (2i + 3j) b_{i,j} = 3 j b_{i+1,j-1} - 2 i b_{i-1,j} + 32 i (i-1) b_{i-2,j+1}.
\]

Computer calculations show that for $i+j \leq 100$ such $p_{i,j} \in \mathbb{Z}$ and $q_{i,j}  \in \mathbb{Z}$ exist that $b_{i,j} = {p_{i,j} \over q_{i,j}}$, $p_{i,j}$ is coprime with $6$ and $q_{i,j}$ is coprime with $6$.

This leads us to the following conjecture:

{\bf Conjecture.}
Let  $a(i, j) = 2^k 3^l s(i,j)$, where $s(i,j) \in \mathbb{Z}$ is coprime with $2$ and $3$. Let
\[
{(4i + 6j + 1)! \over 2^{3 i + 4 j} \, 3^{i+j} \, i! \, j!} = 2^{k_1} 3^{l_1} s_1(i,j), \text{ where } s_1(i,j) \in \mathbb{Z} \text{ is coprime with } 2  \text{ and } 3.
\]
Then $k = k_1$, $l = l_1$.

By the definition $\zeta(u) =(\ln \sigma(u))'$ and
$\wp(u) = -\zeta(u)'$, therefore
\[
\zeta(u) = {\sigma(u)' \over \sigma(u)}, \quad \wp(u) = {\sigma(u)'^2 - \sigma(u)'' \sigma(u) \over \sigma(u)^2}.
\]
Using that $\sigma'(0) = 1$, we see that ${1 \over \zeta(u)}$ and ${1 \over \wp(u)}$ are Hurwitz series over $\mathbb{Z}[{g_2 \over 2}, 2 g_3]$.

\begin{cor} \label{siga}
For any $v$ let $a_2 = \wp(v)$, $a_3 = \wp'(v)$, $a_4 = {g_2 \over 2}$. Then $g_3 = - a_3^2 + 4 a_2^3 - 2 a_2 a_4$.
Thus $\sigma(u) \in H \mathbb{Z}[a_2,a_3,a_4] [[u]]$.
\end{cor}

\subsection{The Baker-Akhiezer function of the elliptic curve.}

The Baker-Akhiezer function plays an important role in the modern theory of integrable systems, see \cite{ref25}. In this work I.~M.~Krichever introduced the addition theorem for this function and demonstrated its important applications.

Consider {\it the Lame equation}
\begin{equation} \label{Lame}
\Phi''(u) - 2 \wp(u) \Phi(u) = \wp(v) \Phi(u).
\end{equation}
The quasiperiodic solutions of (\ref{Lame}) such that $ \underset{u\to
0}{\lim}\left( \Phi(u)-\frac{1}{u} \right)=0$
 are $\Phi(u) = \Phi(u; v)$ and \\ $\Phi_1(u) = \Phi(u; -v)$, where
\begin{equation} \label{BA}
\Phi(u;v) = {\sigma(v-u) \over \sigma(u) \sigma(v)} \exp(\zeta(v) u)
\end{equation}
is {\it the Baker-Akhiezer function}. The periodic properties are
\begin{align}
\Phi(u + 2 \omega_k; v) &= \Phi(u;v) \exp(2 \zeta(v) \omega_k - 2 \eta_k v),
\\
\Phi(u; v + 2 \omega_k) &= \Phi(u;v).
\end{align}
Thus $(\ln \Phi(u))'$ is a doubly periodic meromorphic function of $u$ (see \cite{ref15}):
\begin{equation} \label{lnP'}
(\ln \Phi(u))' = \zeta(u-v) + \zeta(v) - \zeta(u) = {1 \over 2} {\wp'(u) + \wp'(v) \over \wp(u) - \wp(v)}.
\end{equation}
The function $\Phi(u;v)$ considered as a function of $v$ has an exponential singularity in $v=0$.

\begin{rem} \label{KdV} The theory of algebrogeometric solutions of integrable equations like KdV
\begin{equation}
{\partial V \over \partial t} = V''' - 6 V V'
\end{equation}
 started from the work of S.~P.~Novikov \cite{ref35}.
Consider the operators 
\begin{equation} \label{L1}
L_1 = {d^2 \over d u^2} - 2 \wp(u),
\end{equation}
\begin{equation} \label{L2}
L_2 = -2 {d^3 \over d u^3} +6 \wp(u) {d \over d u} +3 \wp'(u).
\end{equation}
The operators $L_1$ and $L_2$ are the Lax pair of the Schrodinger operator $L_1$ with the potential $V(u)=2 \wp(u)$.
The function $2 \wp(u)$ is the solution of the stationary KdV equation. Thus the
operators $L_1$ and $L_2$ commute. 
We have
\[
L_1\Phi(u) = \wp(v) \Phi(u),
\]
\[
L_2\Phi(u) = \wp'(v) \Phi(u).
\]
Thus the fucnction $\Phi(u)$ is the common eigenfunction of the operators $L_1$ and $L_2$ and the pair of eigenvalues
$\big(\wp(v),\wp'(v)\big)$ defines a point on the Weierstrass curve.
\end{rem}

Set $a_2=\wp(v),\; a_3=\wp'(v),\; a_4=\frac{1}{2}g_2$ for the given $v$.

\begin{theorem} \label{lemf0}
In the vicinity of $u=0$ the function $f_0(u) = 1 / \Phi(u)$ is a Hurwitz series over
$\mathbb{Z}[a_2,a_3,a_4]$.
\end{theorem}

The function $f_0(u)$ is regular in the
vicinity of $u=0$. We have
\begin{equation} \label{fuv}
f_0(u) = f_0(u,v) = \sigma(u)\exp \psi(u,v),
\end{equation}
where
\[ \psi(u,v) = \ln \sigma(v)-\ln \sigma(v-u)-\zeta(v)u. \]

The proof of Theorem \ref{lemf0} is based on the following lemma:

\begin{lemma} \label{Lems}
In the vicinity of $u=0$ the function $\psi(u,v)$ is a Hurwitz series over
$\mathbb{Z}[a_2,a_3,a_4]$.
\end{lemma}

Notice that the functions $\wp(v),\; \wp'(v),\; \frac{1}{2}g_2$ are
algebraically independent in the general case, so we can consider $\psi(u,v)$ over a
ring of algebraically independent variables.

{\bf Proof of Lemma \ref{Lems}.}
Teylor decomposition of the function $\psi(u,v)$ at $u=0$
is given by
\begin{equation}
\psi(u,v) = \sum_{k=2}^{\infty}(-1)^k
\Big( -\frac{d^k \ln \sigma(v)}{dv^k} \Big) \frac{u^k}{k!} =
\sum_{k=1}^{\infty}(-1)^{k-1} \wp^{(k-1)}(v) \frac{u^{k+1}}{(k+1)!}.
\end{equation}
Using Weierstrass's uniformization of the elliptic curve (\ref{W}),
we get
\[
\wp''(v)=6 \wp(v)^2-g_2/2.
\]
Thus
\[\wp^{(k)}(v)= p_{k+2}(a_2,a_3,a_4),\; k\geqslant 0, \]
where
\[ \wp^{(0)}(v)=\wp(v)=a_2, \quad \wp^{(1)}(v)=
\frac{\partial}{\partial v}\wp(v)=a_3.\]
It follows that
$p_2(a_2,a_3,a_4)=a_2$ and
\[ p_{k+1}(a_2,a_3,a_4)= \left(a_3 \frac{\partial}{\partial a_2}+
(6a_2^2-a_4)\frac{\partial}{\partial a_3}\right)p_{k}(a_2,a_3,a_4),\;
k\geqslant 2.  \]
Thus
\begin{equation} \label{62}
\psi(u,v) = \sum_{k=2}^{\infty}(-1)^{k} p_{k}(a_2,a_3,a_4)\frac{u^{k}}{k!},
\end{equation}
where $p_{k}(a_2,a_3,a_4)$ is a homogeneous polynomial with integer coefficients of $a_2$, $a_3$, $a_4$.

\begin{rem} \label{rem1} We have $\deg a_2 = -4$, $\deg a_3 = -6$, $\deg a_4 = -8$.
Thus \\ for even $k \geq 0$ we have $p_{k}(a_2,a_3,a_4) = r_k(a_2,a_3^2,a_4)$ for some $r_k(a_2,a_3^2,a_4) \in \mathbb{Z}[a_2,a_3^2,a_4]$.
For even $k \geq 0$ we have $p_{k+3}(a_2,a_3,a_4) = a_3 \, q_k(a_2,a_3^2,a_4)$ for some \\
$q_k(a_2,a_3^2,a_4) \in \mathbb{Z}[a_2,a_3^2,a_4]$.
We have $q_0(a_2,a_3^2,a_4) = 1$.

\end{rem}

\begin{cor}
In the vicinity of $u=0$ the function $\exp\psi(u,v)$ is a Hurwitz series over
$\mathbb{Z}[a_2,a_3,a_4]$.
\end{cor}

By Corollary \ref{siga} $\sigma(u) \in H \mathbb{Z}[a_2,a_3,a_4] [[u]]$. Summarizing this facts, we get the proof of Theorem \ref{lemf0}.

\subsection{The generalized Baker-Akhiezer function.} \text{}

We will need the following functions for the description of the Krichever genus.

Consider the function
\begin{equation} \label{Phihat}
\hat{\Phi}(u) = \Phi(u) \exp(- {\mu_1 \over 2} u),
\end{equation}
where $\Phi(u) = \Phi(u; v)$
is the Baker-Akhiezer function (\ref{BA}),
and the function \\$U_1(u) = U_1(u;v) = - {1 \over 2} {\wp'(u) + \wp'(v) \over \wp(u) - \wp(v)}$.

The function $\hat{\Phi}(u)$ is a solution of the equation
\begin{equation} \label{hP}
\hat{\Phi}''(u) - (2 \wp(u) + \mu_1 U_1(u)) \hat{\Phi}(u) = (\wp(v) + {\mu_1^2 \over 4}) \hat{\Phi}(u).
\end{equation}

It follows from Theorem \ref{lemf0} that
in the vicinity of $u=0$ the function $1 / \hat{\Phi}(u)$ is a Hurwitz series over
$\mathbb{Z}[a_1,a_2,a_3,a_4]$, where $a_1 = {\mu_1 \over 2},\; a_2=\wp(v),\; a_3=\wp'(v),\; a_4=\frac{1}{2}g_2$.

\begin{definition} The generalized Baker-Akhiezer function is $\Psi(u) = \Psi(u; v; \alpha; \mu)$ defined by the formula
\begin{equation} \label{Psi}
\Psi(u) = {\sigma(u +v)^{{1 \over 2} (1 - \alpha)} \sigma(v -u)^{{1 \over 2}
(1 + \alpha)}  \over \sigma(u) \sigma(v)} \exp\left((- {\mu_1 \over 2} +
\alpha \zeta(v)) u\right),
\end{equation}
where $\sigma(u) = \sigma(u; g_2(\mu), g_3(\mu))$, $\zeta(v) = \zeta(v; g_2(\mu), g_3(\mu))$.
\end{definition}

We have
\begin{equation} \label{Psi2}
(\ln \Psi(u))' = {1 \over 2} {\wp'(u) + \alpha \wp'(v) \over \wp(u) -
\wp(v)} - {\mu_1 \over 2}.
\end{equation}

\begin{theorem}
 The function (\ref{Psi}) with $\alpha = {\wp'(w)
\over \wp'(v)}$ satisfies the equation
\begin{equation} \label{P}
\Psi''(u) - (2 \wp(u) + \mu_1 U_1(u) - U_2(u)) \Psi(u) = (\wp(v) + {\mu_1^2 \over 4}) \Psi(u)
\end{equation}
where
$U_2(u) = U_2(u; v, w) = U_1(v; u, w) U_1(v; u, -w)$ and
\begin{equation}
U_1(u; v,w) = -{1 \over 2} {\wp'(u) + \wp'(w) \over \wp(u) - \wp(v)}.
\end{equation}
\end{theorem}

Note that for $w \to v$ or $w \to -v$ (that is $\mu_6 \to 0$) the
equation (\ref{P}) comes to the equation (\ref{hP}), and for
$(\mu_1, \mu_6) \to (0,0)$ to the Lame equation.

The formula holds
\[
\Psi(u; v; \alpha) = \Phi(u;v) \exp(-{\mu_1 \over 2} u) \left({\Phi(u;-v) \over
\Phi(u;v)}\right)^{{1 \over 2}(1 - \alpha)}.
\]

\begin{cor}
The periodic properties are
\begin{equation}
\Psi(u + 2 \omega_k;v;\alpha) = \Psi(u;v;\alpha) \exp(\alpha (2 \zeta(v) \omega_k - 2 \eta_k v) - \mu_1 \omega_k).
\end{equation}
\end{cor}

Let $\alpha = {\wp'(w) \over \wp'(v)}$. Then
\[
\left({\Phi(u;-v) \over \Phi(u;v)}\right)^{{1 \over 2}{\wp'(v) - \wp'(w) \over \wp'(v)}} =
\exp\left( {1 \over 2} (\wp'(v) - \wp'(w)) {\ln(\sigma(u+v)) - \ln(\sigma(v-u)) - 2 \zeta(v) u \over \wp'(v)} \right).
\]

Let $\omega_k$, $k = 1,2,3$ be the half-periods of the elliptic curve in the standard Weierstrass form (\ref{Wei}), where $\omega_1 + \omega_2 + \omega_3 = 0$.

We have
\[
\Phi(u; \omega_k) = \Phi(u; - \omega_k) = {\sigma(\omega_k - u)
\over \sigma(u) \sigma(\omega_k)} \exp(\eta_k u).
\]

Thus $\Phi(u; \omega_k)^2 = \wp(u) - e_k$, where $e_k = \wp(\omega_k)$.

It is wellknown that $\Phi(u; \omega_k) = {1 \over sn(u; \delta, \varepsilon)}$, where
 $\delta = - {3 \over 2} e_k$, $\varepsilon = 3 e_k^2 - {g_2 \over 4}$.

Note that $\wp'(v) \to 0$ if and only if $v \to \omega_k$, $k = 1,2,3$.

\begin{lemma}
We have
\[
\lim_{v \to \omega_k} \Psi(u) =\Phi(u;\omega_k) \exp\big(-{\mu_1
\over 2} u+W(u)\big),\] where $W(u)=\left( {1 \over 2} {e_k -
\wp'(w) \over 6 e_k^2 - {g_{2} \over 2}} (\zeta(u+\omega_k) - \zeta(\omega_k-u) + 2 e_k u) \right)$.
\end{lemma}

{\bf Proof.}
We have
\begin{multline*}
\lim\limits_{v \to \omega_k} {\ln\big(\sigma(u+v)\big)
- \ln\big(\sigma(v-u)\big) - 2 \zeta(v) u \over \wp'(v)} =\\= \lim_{v \to \omega_k} {\zeta(u+v) - \zeta(v-u) + 2 \wp(v) u \over \wp''(v)} =
{\zeta(u+\omega_k) - \zeta(\omega_k-u) + 2 e_k u \over 6 e_k^2 - {g_2 \over 2}}
\end{multline*}
which is a meromorphic function without pole in zero.

Let $\hat{\Psi}$ be the function from the paper \cite{ref33}
\[
\hat{\Psi}(x; z, \eta) = {\sigma(z+x+\eta) \over \sigma(z+\eta) \sigma(x)} \left[{\sigma(z-\eta)\over \sigma(z+\eta)}\right]^{x/(2 \eta)}.
\]

\begin{theorem} \label{thisth}
\[
\hat{\Psi}(u; z, \eta) = \Psi(u; v; -1) \] \[\text{ for } v = z + \eta, \quad - { \mu_1 \over 2} = \zeta(z+\eta) + {1 \over 2 \eta} \ln \left[{\sigma(z-\eta)\over \sigma(z+\eta)}\right].
\]
\end{theorem}

{\bf Proof.} $\lim_{u \to 0} u \Psi(u) = 1 = \lim_{u \to 0} u \hat{\Psi}(u)$.
The logarithm derivatives of this functions are equal for the given $v$, $\mu_1$.

We will get the results similar to the Remark \ref{KdV} for the function $\Psi(u)$ in our following works.

\section{\bf The general elliptic genus.}

{\it The general elliptic genus} is the Hirzebruch
genus $L_f$, where $f$ is the exponential of the general elliptic
formal group law $F_{\mu}(t_1, t_2)$.

\begin{theorem} (Integrality of the Hirzebruch genus.)
The general elliptic formal group law $F_{\mu}(t_1, t_2)$ defines a
5-parametric family of $E$-integer Hirzebruch genera.
\end{theorem}
{\bf Proof.} The proof follows from the fact that the formal group of geomertic cobordisms (\ref{Fcc}) is universal and the general elliptic formal group law is defined over the ring \\ $E = \mathbb{Z}[\mu_1, \mu_2, \mu_3, \mu_4, \mu_6]$.

\begin{cor}
Let $(\mu_4, \mu_6) = (0,0)$. Then the corresponding formal group law is universal over the set of formal group laws over graduate rings $A$ that are multiplicatively generated by $a_k$, $\deg a_k = - 2 k$, $k = 1,2,3$.
\end{cor}
{\bf Proof.} The proof follows from the fact that the ring of coefficients of the formal group law generates all the ring.

\begin{cor}
 Let $\delta = \mu_2$,\; $\varepsilon = \mu_2^2 - 4 \mu_4$. The Hirzebruch genus $L_f$ with the exponential $f(u) =
sn(u) $ such that
\[
f'(u)^2 = 1 - 2 \delta f(u)^2 + \varepsilon f(u)^4
\]
defines an $A$-integer Hirzebruch genus, where $A = \mathbb{Z}[\mu_2, \mu_4]$.
\end{cor}

\section{\bf The general Krichever genus.}

\subsection{The Krichever genus.}

Let $f_0(u) = {1 \over \Phi(u)}$, where $\Phi(u)$ is the Baker-Akhiezer function (\ref{BA}).

In the work \cite{ref31} Krichever introduced the Hirzebruch genus defined by the function $f_0(u)$ and it was shown
that it obtains the remarkable property of rigidity on SU-manifolds (Calabi–Yau manifolds) with the action of a circle $S^1$.

Let $a_2 = \wp(v)$, $a_3 = \wp'(v)$, $a_4 = {1 \over 2} g_2$.

The Hirzebruch genus
\[
 L_{Kr}: \Omega_U \longrightarrow \mathbb{Q}[a_1, a_2, a_3, a_4],
\]
defined by the series $f_{Kr}(u)= f_0(u) \exp(a_1 u)$, is called {\it the Krichever genus}.
See \cite{ref31}.

Consider the transform
\begin{equation}\label{Pu}
T\big(f(u)\big) = \frac{f(u)}{f'(u)}.
\end{equation}
It brings Hurwitz series $f(u)$ such that $f(0)=0,\; f'(0)=1$ to Hurwitz series with the same property.

We have
\begin{equation} \label{G}
\frac{d}{du} \ln f_{Kr}(u) = a_1- \frac{1}{2}\,\frac{\wp'(u)+\wp'(v)}{\wp(u)-\wp(v)}.
\end{equation}

Comparing the formulas (\ref{exp2}) and (\ref{G}) we obtain the following result:

\begin{lemma} The transform {\rm(\ref{Pu})} brings $f_{Kr}$ to the exponential of the ellipic formal group law, where $\wp'(v)$ = $\wp'(w)$ and $a_1 = {\mu_1 \over 2}$. Thus we obtain $\mu_1=2a_1, \; \mu_2=3a_2-a_1^2,$ $\mu_3=-a_3, \; \mu_4 = 3 a_2^2 - a_1 a_3 - {1 \over 2} a_4, \; \mu_6 = 0$.
\end{lemma}

\begin{rem}
For $a_1 = {\mu_1 \over 2}$ we have by definition $f_{Kr} = {1 \over \hat{\Phi}(u)}$.
\end{rem}

\begin{cor} \label{corPhi2}
Let $f(u)$ be the exponential of the elliptic formal group law $F_{\mu}(t_1, t_2)$ where $\mu = (\mu_1, \mu_2, \mu_3, \mu_4, 0)$. Let $\wp(v) = {1 \over 12} (4 \mu_2 + \mu_1^2)$, $\wp'(v) = - \mu_3$.
Then
\[(\ln \hat{\Phi}(u))' = - {1 \over f(u)} .\]
\end{cor}

For $a_1 = 0$ we obtain the result for the Baker-Akhiezer function (\ref{BA}):
\begin{cor} \label{corPhi}
Let $f(u)$ be the exponential of the elliptic formal group law $F_{\mu}(t_1, t_2)$ where $\mu = (0, \mu_2, \mu_3, \mu_4, 0)$. Let $\wp(v) = {1 \over 3}  \mu_2$, $\wp'(v) = - \mu_3$.
Then
\[{\partial \over \partial u} \ln \Phi(u; v) = - {1 \over f(u)} .\]
\end{cor}

\subsection{The general Krichever genus.}

The Corollary \ref{corPhi} can be reformulated in the following way: The Baker-Akhiezer function $\Phi(u; v)$ for the proper $v$ is a solution of the equation
\[
\Phi'(u) + {1 \over f(u)} \Phi(u) = 0,
\]
where $f(u)$ is the exponential of the elliptic formal group law with the parametra $\mu = (0, \mu_2, \mu_3, \mu_4, 0)$. By the Corollary \ref{corPhi2} the function $\hat{\Phi}(u)$ is the solution of the same equation for the elliptic formal group law with the parametra $\mu = (\mu_1, \mu_2, \mu_3, \mu_4, 0)$. The function $f_{Kr} = {1 \over \hat{\Phi}(u)}$ defines the Krichever genus. Let us define the general Krichever genus in the following way:

{\it The general Krichever genus} is the Hirzebruch genus $L_\phi$, where $\phi(u) = {1 \over \Psi(u)}$, and $\Psi(u)$ the solution of
the equation
\[
\Psi' + {1 \over f(u)} \Psi = 0
\]
such that $\lim\limits_{u \to 0} \big(\Psi(u) - {1 \over u}\big) =\,$const,
where $f(u)$ is the exponential of the general elliptic formal group law.

Comparing (\ref{Psi2}) and (\ref{exp2}), we come to the following theorem:

\begin{theorem} \label{th30}
The exponential of the formal group law corresponding to the general Krichever genus has the form
\[
{1 \over \Psi(u)} = \Phi(u;v)^{-1} \exp\left({\mu_1 \over 2} u\right)
\left({\Phi(u;v) \over \Phi(u;-v)}\right)^{{1 \over 2}(1 - \alpha)},
\quad \text{where} \]
where $\wp(v) = {1 \over 12} (4 \mu_2 + \mu_1^2)$, $\wp'(w) = - \mu_3$ and $\alpha = {\wp'(w) \over \wp'(v)}$.
\end{theorem}

\begin{rem}
The function $\Psi(u)$ conclides with the one defined by (\ref{Psi}).
\end{rem}

\begin{cor}
The general Krichever genus becomes the Krichever genus for $v=w$.
\end{cor}

\begin{lemma} \label{HirwPsi}
In the vicinity of $u = 0$ the exponential of the general Krichever genus
\[
{1 \over \Psi(u)} = u + \sum \Psi_k {u^{k+1} \over (k+1)!},
\]
is a Hurwitz series over $\mathbb{Z}[a_1, a_2, a_3, a_4, a_6]$, where
$a_1 = {\mu_1 \over 2}$, $a_2 = \wp(v) = {1 \over 12} (4 \mu_2 + \mu_1^2)$,
$a_3 = \wp'(w) = - \mu_3$, $a_4 = {1 \over 2} g_2(\mu)$, $a_6 = 4 \mu_6 + \mu_3^2$.
\end{lemma}

{\bf Proof.}
We have
\begin{equation} \label{1/Psi}
{1 \over \Psi(u)} = \sigma(u) \exp({\mu_1 \over 2} u) \exp(\psi(u,v)),
\end{equation}
where
\[ \psi(u,v) = \ln \sigma(v) - {1 \over 2} (1 - \alpha) \ln \sigma(v+u) - {1 \over 2} (1 + \alpha) \ln \sigma(v-u) -
\alpha \zeta(v) u. \]
Teylor decomposition of the function $\psi(u,v)$ at $u=0$
is given by
\begin{equation}
\psi(u,v) =
\sum_{k=1}^{\infty} {1 \over 2} ( (1 - \alpha) + (-1)^{k-1} (1 + \alpha)) \wp^{(k-1)}(v) \frac{u^{k+1}}{(k+1)!}.
\end{equation}
Using the Remark \ref{rem1} and the notation $b_3 = \wp'(v)$ we obtain
\begin{equation}
\psi(u,v) = \sum_{n=0}^{\infty} r_{2n+2}(a_2,b_3^2,a_4) \frac{u^{2n+2}}{(2n+2)!} - (\alpha \, b_3)
\sum_{n=0}^{\infty} q_{2n}(a_2,b_3^2,a_4) \frac{u^{2n+3}}{(2n+3)!}.
\end{equation}
Thus $\psi(u,v)$ is a Hirwitz series over $a_2$, $a_6 = b_3^2$, $a_4$ and $a_3 = \alpha \, b_3$.\\
We have $\sigma(u) \in H \mathbb{Z}[{g_2 \over 2}, 2 g_3] [[u]]$ and thus $\sigma(u) \in H \mathbb{Z}[a_2, b_3^2, a_4] [[u]]$.\\
Thus ${1 \over \Psi(u)} \in H\mathbb{Z}[a_1, a_2, a_3, a_4, a_6]$.

\subsection{The formal group law for the Krichever genus.} \text{}

The addition theorem, characterising the Krichever genus, was introduced in \cite{ref24}.
The universal properties of this genus were described in \cite{ref32}. Unfortunately, the proof of theorem 6.23 of this work contains inaccuracies.

Following \cite{ref25}, we can write the addition theorem for the function $f_0(u) = 1 / \Phi(u, v)$ in the form
\begin{equation} \label{179}
f_0(u+v) = {f_0(u)^2 {f_0(v)\over -f_0(-v)} - f_0(v)^2 {f_0(u)\over -f_0(-u)} \over f_0(u) f_0'(v) - f_0(v) f_0'(u)}.
\end{equation}

Notice that if the function $f_0(u)$ gives a solution of the equation (\ref{179}), then the product $\exp(cu) \, f_0(u)$ also gives a solution for any constant $c$.

The followig theorem characterises the Krichever genus $f_{Kr}$ in terms of addition theorems. All the series considered in the theorem below are over $A \otimes \mathbb{Q}$.

\begin{theorem}{\rm(A version of theorem 1 from \cite{ref24})} \label{3.1}
The function $f(u)$ such that \\
$f(0)=0,\; f'(0)=1$ has an addition theorem of the form
\begin{equation}\label{1'''}
f(u+v)=
\frac{f(u)^2\xi_1(v)-f(v)^2\xi_1(u)}{f(u)\xi_2(v)-f(v)\xi_2(u)}
\end{equation}
(for some series $\xi_1(u)$ and $\xi_2(u)$ such that $\xi_1(0)=\xi_2(0)=1$)
if and only if $f(u) = f_{Kr}(u)$ is the Krichever genus.
\end{theorem}

We will give the proof in a few steps:

\begin{rems}\text{} \label{rem50}
\begin{enumerate}
\item Let $f(u)$ satisfy the equation (\ref{1'''}) with some $\xi_1(u)$
and $\xi_2(u)$. Then the series $f(u)\exp (a_1u)$ also satisfies the equation (\ref{1'''}), where the series $\xi_1(u)$ and $\xi_2(u)$ are replaced by the series $\xi_1(u)\exp (2 a_1u)$ and $\xi_2(u)\exp (a_1u)$ .
\item Let $f(u)$ satisfy the equation (\ref{1'''}) with some $\xi_1(u)$
and $\xi_2(u)$. Then the same series $f(u)$ satisfies the equation
(\ref{1'''}), where the series $\xi_1(u)$ and $\xi_2(u)$ are replaced by the series
$\xi_1(u)+\gamma_2 f(u)^2$ and $\xi_2(u)+\gamma_1 f(u)$.
\end{enumerate}
\end{rems}

\begin{lemma}\label{3.2}
Let the function $\widetilde f(u)$ such that $\widetilde f(0)=0,\; \widetilde f'(0)=1$ be a solution of the addition theorem of the form (\ref{1'''}).
Then $\widetilde f(u)=
f(u)\exp (a_1u)$, where $f(u)$ is the solution of the differential equation
\begin{equation} \label{2'''}
\left( f'''+2 a_2 f'- a_3 f \right) f-3 f'' f'=0,
\end{equation}
with initial conditions $f(0)=0,\; f'(0)=1,\; f''(0)=0$.
Here $a_1,\,a_2,\,a_3 \in A$.
\end{lemma}

{\bf Proof.}
Taking into account the Remarks \ref{rem50}, it is sufficient to prove the lemma with the following initial conditions:
\[ f(0)=f''(0)=0,\; f'(0)=1;\quad \xi_1(0)=\xi_2(0)=1,\; \xi_1''(0)=\xi_2'(0)=0. \]
We have
\begin{equation} \label{3}
f(u+v)\left[ f(u)\xi_2(v)-f(v)\xi_2(u)
\right]=f(u)^2\xi_1(v)-f(v)^2\xi_1(u).
\end{equation}
Set $f(u)=f,\; \xi_1(u)=\xi_1,\; \xi_2(u)=\xi_2$.
The series
$f(v),\; \xi_1(v),\; \xi_2(v)$ up to $v^4$ have the following form:
\[
f(u+v) \thickapprox f+f'v+f''\frac{v^2}{2}+f'''\frac{v^3}{3!},
\]
\[
f(v) \thickapprox v+f_2\frac{v^3}{3!};\quad
\xi_1(v) \thickapprox 1+\xi_{1,1}v+\xi_{1,3}\frac{v^3}{3!};\quad
\xi_2(v) \thickapprox
1+\xi_{2,2}\frac{v^2}{2}+\xi_{2,3}\frac{v^3}{3!}\,.
\]
Substituing into (\ref{3}):
\begin{multline*}
\left( f+f'v+f''\frac{v^2}{2}+f'''\frac{v^3}{3!} \right)\left[
f\Big(1+\xi_{2,2}\frac{v^2}{2}+\xi_{2,3}\frac{v^3}{3!}\Big)-\Big(v+f_2\frac{v^2}{3!}
\Big)\xi_2\right]=\\
=f\left[
f+f'v+f''\frac{v^2}{2}+f'''\frac{v^3}{6}+\xi_{2,2}f\frac{v^2}{2}+\xi_{2,2}f'\frac{v^3}{2}
+\xi_{2,3}f\frac{v^3}{3!} \right]-\Big(
fv+f'v^2+f''\frac{v^3}{2}+f_2f\frac{v^3}{6} \Big)\xi_2\,.
\end{multline*}
On the other hand
\[ f^2\Big(1+\xi_{1,1}v+\xi_{1,3}\frac{v^3}{6}\Big)- \Big(v+f_2\frac{v^2}{6}
\Big)^2\xi_1 = f^2\Big(1+\xi_{1,1}v+\xi_{1,3}\frac{v^3}{6}\Big)-
\xi_1 v^2. \] Comparing the coefficients at the corresponding degrees of $v$, we get:

At $v$ we have
\[
ff'-f\xi_2 = \xi_{1,1}f^2.
\]
Therefore $\xi_2 = f'-\xi_{1,1}f$. We have:
$\xi_2'(0)=f''(0)-\xi_{1,1}f'(0)$. Thus, if
$f''(0)=\xi_2'(0)=0$ and $f'(0)=1$, then $\xi_{1,1}=0$ and $\xi_2(u) =
f'(u)$.

At $v^2$ we have
\[ \frac{1}{2}ff''+\frac{1}{2}\xi_{2,2}f^2-f'\xi_2 = -\xi_1. \]
Using $\xi_2(u) = f'(u)$, we get:
\[ \xi_1(u) = f'(u)^2-\frac{1}{2}ff''-\frac{1}{2}f_2f^2. \]
Thus, if $f''(0)=0$ and $f'(0)=1$, then the equation (\ref{1'''})
is equivalent to the equation
\begin{equation}\label{4}
f(u+v)=f(u)f'(v)+f(v)f'(u)-
\frac{1}{2}f(u)f(v)\frac{f(u)f''(v)-f(v)f''(u)}{f(u)f'(v)-f(v)f'(u)}\,.
\end{equation}
Set $\psi(u)=\big(\ln f(u)\big)'$. Then the equation (\ref{4})
can be presented in the form
\[ f(u+v)=\frac{1}{2}f(u)f(v)\left[
\psi(u)+\psi(v)-\frac{\psi'(u)-\psi'(v)}{\psi(u)-\psi(v)}\right].
\]
Thus,
\begin{equation}
f(u+v)=\frac{1}{2}f(u)f(v)\left[
\psi(u)+\psi(v)-\partial_+\ln\big(\psi(u)-\psi(v)\big)\right].
\end{equation}

At $v^3$ we have
\[ \frac{1}{6}ff'''+\frac{1}{2}\xi_{2,2}ff' +\frac{1}{6} \xi_{2,3}f^2
-\frac{1}{2}f''\xi_{2}-\frac{1}{6} f_2f\xi_2 =
\frac{1}{6}\xi_{1,3}f^2. \]
Because  $\xi_2(u) = f'(u)$ and thus $\xi_{2,2}=f_2$, we get
\[
\left[ f'''+2f_2f'+(\xi_{2,3}-\xi_{1,3})f \right]f-3f''f'=0.
\]
Setting $a_2=f_2$ and $a_3=\xi_{2,3}-\xi_{1,3}$, we get the proof of the lemma.

\begin{cor}
Let the function $\widetilde f(u)$ such that $\widetilde f(0)=0,\; \widetilde f'(0)=1$ be a solution of the addition theorem of the form (\ref{1'''}).
 Then it has an addition theorem of the form
\begin{equation}
 f(u+v)=\frac{1}{2}f(u)f(v)\left[
 \psi(u)+\psi(v)-\frac{\psi'(u)-\psi'(v)}{\psi(u)-\psi(v)}\right]
\end{equation}
where $\psi(u)=\frac{f'(u)}{f(u)}=\frac{1}{u}+\psi_0(u)$, $\psi_0(u) \in A \otimes \mathbb{Q}[[u]]$, and $\psi(u)$ satisfies the differential equation
\begin{equation}
(\psi')^2 = \psi^4-2 a_2\psi^2+ a_3 \psi- a_4.
\end{equation}
\end{cor}

{\bf Proof.}
Using the formulae
\[ \frac{f''}{f} = \psi'+\psi^2,\quad \frac{f'''}{f} = \psi''+3\psi'\psi+\psi^3, \]
we get from (\ref{2'''}) the equation
\[ (\psi''+3\psi'\psi+\psi^3+2 a \psi- a_3)-3(\psi'+\psi^2)\psi=0. \]
Thus
\begin{equation} \label{8}
\psi''-2\psi^3+2 a_2\psi- a_3 = 0.
\end{equation}
Multiplying the equation (\ref{8}) by $2\psi'$ and integrating, we come to the equation
\[ (\psi')^2 = \psi^4 - 2 a_2\psi^2 + a_3\psi - a_4. \]

{\bf Last step of the proof of theorem \ref{3.1}.}
Consider the function
\[ f(u)= {1 \over \Phi(u; w)} =  \frac{\sigma(u)\sigma(w)}{\sigma(w-u)}\exp(-\zeta(w)u). \]
We have $ f(u)=\sigma(u)\exp\psi(u,w)$ with $\psi(0,w) = 0$, $\psi'(0,w) = 0$.\\
Thus $f(0)=f''(0)=0,\; f'(0)=1$.

The given function has an addition theorem (\ref{179})
of the form (\ref{1'''}), where $\xi_1(u)=- \frac{f(u)}{f(-u)},\;
\xi_2(u)=f'(u)$. Thus the function $f(u)$ is the solution of the equation (\ref{2'''}) with the given initial conditions.
Using the uniqueness of the solution of this equation, we get the proof.

\begin{rem}
In the proof of the lemma \ref{3.2} we have obtained the formula
\[ \xi_1(u)=(f')^2-\frac{1}{2}ff''-\frac{1}{2}f_2f^2. \]
Thus, for the given function $f(u)$ we have
\[f(-u) = -{f(u) \over (f'(u))^2- \frac{1}{2}f(u) f''(u) -\frac{1}{2} f_2 f(u)^2}.\]
Therefore in the case of an odd function $f(u)$ we get the equation
\[ (f')^2- \frac{1}{2}ff''-\frac{1}{2} f_2 f^2=1 \quad \text{ with the initial conditions } \quad f(0)=0,\; f'(0)=1, \]
and the solution $f(u)=sn(u)$.
\end{rem}

\section{\bf Appendix and Applications.}

The following examples are related to the authomorphisms of the elliptic formal group laws.

\begin{example} \label{ex1}
In the notations of section 2.9, let $n=2$ and $\alpha = -1$. Then $\mu_1 = \mu_3 = 0$.

In this case we obtain an elliptic curve in Tate coordinates (\ref{Ats})
\[
s = t^3 + \mu_2 t^2 s + \mu_4 t s^2 + \mu_6 s^3.
\]

Set $s=tv$. Then
$$\frac{v}{1+\mu_2v+\mu_4v^2+\mu_6v^3} = \tau$$ where
$\tau=t^2$.\; Set $1+\mu_2z+\mu_4z^2+\mu_6z^3 =
(1+\gamma_1z)(1+\gamma_2z)(1+\gamma_3z)$. Using the classical
Lagrange inversion formula, we obtain
\begin{align*}
v(\tau) &= -\frac{1}{2\pi i}\oint_{|z|=\varepsilon}\ln\left[
1-\frac{\tau}{z}(1+\gamma_1z)(1+\gamma_2z)(1+\gamma_3z)\right]dz
=\\
&= \sum_{n=1}^{\infty}\frac{\tau^n}{n}\,\frac{1}{2\pi
i}\oint_{|z|=\varepsilon}
\frac{(1+\gamma_1z)^n(1+\gamma_2z)^n(1+\gamma_3z)^n}{z^n}\,dz =
 \\
&= \sum_{n\geqslant 1}\sum_{j_1+j_2+j_3=n-1} \binom{n}{j_1}
\binom{n}{j_2} \binom{n}{j_3}\gamma_1^{j_1} \gamma_2^{j_2}
\gamma_3^{j_3} \frac{\tau^n}{n}\,.
\end{align*}
Therefore we have
\begin{equation} \label{binom}
s(t) = t \sum_{n\geqslant 1}\sum_{j_1+j_2+j_3=n-1} \binom{n}{j_1}
\binom{n}{j_2} \binom{n}{j_3}\gamma_1^{j_1} \gamma_2^{j_2}
\gamma_3^{j_3} \frac{t^{2n}}{n}
\end{equation}
where $\gamma_1+\gamma_2+\gamma_3=\mu_2,\;
\gamma_1\gamma_2+\gamma_1\gamma_3+\gamma_2\gamma_3=\mu_4,\;
\gamma_1\gamma_2\gamma_3=\mu_6$.

Formula (\ref{246}) gives the elliptic formal group law
\[
F_{\mu}(t_1, t_2) = t_1 + t_2 + b {(\mu_2 + 2 \mu_4 m + 3 \mu_6 m^2) \over (1 + \mu_2 m + \mu_4 m^2 + \mu_6 m^3)}.
\]
Formula (\ref{geneq}) gives the equation on the exponential
\begin{multline} \label{fcub}
\mu_6 f'(u)^3 + (3 \mu_6 + (\mu_4^2 - 3 \mu_2 \mu_6) f(u)^2) f'(u)^2 + (27 \mu_6^2 - \mu_2^2 \mu_4^2 + 4 \mu_2^3 \mu_6 - 18 \mu_2 \mu_4 \mu_6 + 4 \mu_4^3) f(u)^6 + \\ + (18 \mu_4 \mu_6 - 12 \mu_2^2 \mu_6 + 2 \mu_2 \mu_4^2) f(u)^4 + (12 \mu_2 \mu_6 - \mu_4^2) f(u)^2 - 4 \mu_6 = 0.
\end{multline}

\begin{lemma}
The exponential $f(u)$ of an elliptic formal group law is an odd function if and only if $\mu_1 = \mu_3 = 0$.
\end{lemma}

{\bf Proof.} Let $\mu_1 = \mu_3 = 0$. Then it follows form (\ref{Ats}) that $s(t)$ is an odd function of $t$, thus $b(t, -t) = 0$. It follows from (\ref{246}) that $F_{\mu}(t, - t) = 0$, thus $g(t) = - g(-t)$ and $f(t) = - f(-t)$. The inverse follows from (\ref{f'}) and the series decomposition of the right part of (\ref{dFdt}).

\begin{cor}
In the considered case the exponential $f(u) \in \mathbb{Z}_{(2)}[\mu_2, \mu_4, \mu_6][[u]]$ and the logarithm $g(t) \in \mathbb{Z}_{(2)}[\mu_2, \mu_4, \mu_6][[t]]$. Thus the exponential gives a strong isomorphism over $\mathbb{Z}_{(2)}[\mu_2, \mu_4, \mu_6]$ of the formal group law $F_{\mu}(t_1, t_2)$ given by (\ref{246}) and the linear group $L$.
\end{cor}
\end{example}

\begin{example} \label{ex2}
Let $n=3$ and let $\alpha$ be a root of $1$ of order $3$. Then we obtain
{\it the equianharmonic case} $\mu_1 = \mu_2 = \mu_4 = 0$.

In this case we obtain an elliptic curve in Tate coordinates (\ref{Ats})
\[
s =  t^3 + \mu_3 s^2 + \mu_6 s^3 \quad \text{ with } \quad \Delta =  - 27 (4 \mu_6 + \mu_3^2)^2.
\]

The elliptic formal group law (\ref{second}) takes the form
\[
F_{\mu}(t_1, t_2) = {  (t_1 + t_2) (1 + \mu_6 m^3) + \mu_3 m^2 + 3 \mu_6 m^2 b \over  (1 + \mu_6 m^3) (1 - \mu_3 b) - {b m \over p} \mu_3 (1 - \mu_3 b - \mu_6 b^2) }.
\]

Formula (\ref{geneq}) gives an equation on its exponential $f(u)$:
\begin{equation} \label{f36}
\mu_6 \left[ f'(u)^3 + 3 f'(u)^2 + 27 \mu_6 f(u)^6 + 18 \mu_3 f(u)^3 - 4 \right] = - \mu_3^2 \left[ f'(u)^2 + 4 \mu_3 f(u)^3 - 1 \right].
\end{equation}

\end{example}

\begin{example} \label{ex3}
Let $n=4$ and let $\alpha$ be a root of $1$ of order $4$. Then we obtain
{\it the Lemniscate case} $\mu_1 = \mu_2 = \mu_3 = \mu_6 = 0$.

In this case we obtain an elliptic curve in Tate coordinates (\ref{Ats})
\[
s =  t^3 + \mu_4 t s^2, \quad \text{ with } \quad \Delta = - 64 \mu_4^3.
\]
Let $v = \mu_4 t s$ and $\tau = \mu_4 t^4$. The equation becomes
\[
v^2 - v + \tau = 0, \quad \text{thus} \quad v(\tau) = {1 \over 2} (1 - \sqrt{1 - 4 \tau}).
\]

Formula (\ref{F1}) gives the elliptic formal group law
\[
F_{\mu}(t_1, t_2) = t_1 + t_2 + {2 \mu_4 b m \over 1 + \mu_4 m^2}.
\]

Formula (\ref{geneq}) gives an equation on its exponential $f(u)$:
\[
f(u)' = \sqrt{1 - 4 \mu_4 f(u)^4}.
\]
It implies from (\ref{fg}) that the function $f(u) = {- 2 \wp(u; 4 \mu_4, 0) \over \wp'(u; 4 \mu_4, 0)}$ is a solution of this equation and the exponential of the formal group $F_{\mu}$.

Notice that we can choose the half-periods $\omega_1$ and $\omega_2$ of the function $\wp(u; 4 \mu_4, 0)$ such that $\omega_2 = i \omega_1$.

Now let us use equation (\ref{generators}) to find the image of $\mathcal{A}_{(2)} \to \mathbb{Z}[\mu_4]$.
We have
\[
F_{\mu}(t, t) =  2 t {1 - 2 \mu_4 t s \over 1 + 4 \mu_4 t^4},
\qquad
{\partial \over \partial t_2} F_{\mu}(t, t_2) |_{t_2 = 0} = {2 t^3 \over s} - 1.
\]
The function $Cat(\tau) = {1 \over 2 \tau} (1 - \sqrt{1 - 4 \tau})$ is the generating function of the Catalan numbers, i.e. $Cat(\tau) = \sum_{n \geqslant 0} C_n \tau^n$, where $C_n = {1 \over n+1} \binom{2n}{n}$.
Using our notations we obtain $v(\tau) =$~$\tau Cat(\tau)$, and $s = t^3 Cat(\mu_4 t^4)$. Thus
\[
s(t) = \sum_{n \geqslant 0} C_n \mu_4^{n} t^{4n+3}.
\]
On the other hand it follows from (\ref{binom}) that $C_n = {1 \over n+1} \binom{2n}{n} = (-1)^n \frac{1}{2n+1} \sum_{j=0}^{2n} (-1)^{j} \binom{2n+1}{j}
\binom{2n+1}{j+1}.$

Let $a_i$ for $i = 1,2,3,...$ be the generators of $\mathcal{A}_{(2)}$.
For $\phi: \mathcal{A}_{(2)} \to \mathbb{Z}[\mu_4]$ we have $\phi(a_i) = 0$ for $i \ne 4 k$, $k = 1,2,3,... \,$.
Thus
\[
{\partial \over \partial t_2} F_{\mu}(t, t_2) |_{t_2 = 0} = 1 + \sum_{k \geqslant 1} \phi(a_{4k}) t^{4 k}.
\]
Summarizing this formulas, we get
\[
(2 + \sum_{k \geqslant 1} \phi(a_{4k}) t^{4 k}) (\sum_{n \geqslant 0} C_n \mu_4^{n} t^{4n}) = 2.
\]
Thus $\phi(a_{4 k})$ are given by the system of formulas
\[
\sum_{q=0}^m \phi(a_{4(m-q)}) C_{q} \mu_4^{q} = 0, \quad \text{where} \quad \phi(a_{0}) = 2, \quad m \geq 1.
\]
For $m = 1$ we get $\phi(a_{4}) = - 2 \mu_4$. We get the relations between $\phi(a_{4q})$:
\[
\sum_{q=0}^m (-1)^q 2^{m-q-r(m)} C_{q} \phi(a_{4(m-q)}) \phi(a_{4})^{q} = 0, \quad r(m) = \text{gcd}(2 C_m, 2^{m-q} C_{q}), \quad m \geq 1.
\]

Thus the image of $\mathcal{A}_{(2)} \to \mathbb{Z}[\mu_4]$ the generators can be chosen as $\alpha_m = \phi(a_{4m})$, and the image will be
\[
\mathbb{Z}[\alpha_m] / J
\]
where $J$ is generated by $(-1)^{m} 2^{1-r(m)} C_{m} \alpha_1^{m} + \sum_{q=1}^m (-1)^{m-q} 2^{q-r(m)} C_{m-q} \alpha_{q} \alpha_1^{m-q}$.

Explicitly, this equations for $m = 2,3,4$ will be
$2 \alpha_{2} = - \alpha_1^{2}$,
$4 \alpha_{3} = 3 \alpha_1^{3} + 2 \alpha_{2} \alpha_1 = 2 \alpha_1^{3}$, and
$8 \alpha_{4} = - 9 \alpha_1^{4} -  4 \alpha_{2} \alpha_1^{2} + 4 \alpha_{3} \alpha_1 = - 5 \alpha_1^{4}$.

\end{example}

\begin{example} \label{ex4}
Let $n=4$ and let $\alpha$ be a root of $1$ of order $6$. Then $\mu_1 = \mu_2 = \mu_3 = \mu_4 = 0$.

In this case we obtain an elliptic curve in Tate coordinates (\ref{Ats})
\[
s = t^3 + \mu_6 s^3, \quad \text{ with } \quad
\Delta =   - 432 \mu_6^2.
\]
Formula (\ref{F1}) gives the elliptic formal group law
\[
F_{\mu}(t_1, t_2) = t_1 + t_2 + { 3 \mu_6 b m^2 \over 1 + \mu_6 m^3}.
\]
Formula (\ref{geneq}) gives a differential equation on its exponential $f(u)$:
\[
27 \mu_6  f(u)^6  = (1 - f'(u)) (2 + f'(u))^2.
\]
It implies from (\ref{fg}) that the function $f(u) = {- 2 \wp(u; 0, 4 \mu_6) \over \wp'(u; 0, 4 \mu_6)}$ is a solution of this equation and the exponential of the formal group.

Notice that we can choose the half-periods of the function $\wp(u; 0, 4 \mu_6)$ $\omega_1$ and $\omega_2$ such that $\omega_2 = {1 + i \sqrt{3} \over 2} \omega_1$.

\end{example}

Now let us consider some examples that are not related to the authomorphisms of the elliptic formal group laws directly.

\begin{example} \label{ex5} Let $\mu_3 = \mu_4 = \mu_6 = 0$.

In this case we obtain an elliptic curve in Tate coordinates (\ref{Ats})
\[
(1 - \mu_1 t - \mu_2 t^2) s = t^3 \quad \text{ with } \quad  \Delta = 0.
\]
The elliptic formal group law (\ref{123}) takes the form
\[
F_{\mu}(t_1, t_2) =  { t_1 + t_2 - \mu_1 t_1 t_2 \over 1 + \mu_2 t_1 t_2}.
\]
The exponential is a solution of the equation (\ref{Ricatti}):
\[
f(u)' = 1 - \mu_1 f(u) - \mu_2 f(u)^2.
\]
Let $\alpha + \beta = \mu_1$, $\alpha \beta = - \mu_2$. The solution of this equation is
\[
f(u) = {e^{\alpha u} - e^{\beta u} \over \alpha e^{\alpha u} - \beta e^{\beta u}}.
\]
We have (see \cite{ref5}):
\[
{e^{\alpha u} - e^{\beta u} \over \alpha e^{\alpha u} - \beta e^{\beta u}} = \sum_{n=0}^\infty \sum_{k=0}^n (-1)^n A_{n,k} \alpha^n \beta^{n-k} {u^{n+1} \over (n+1)!}.
\]
Here $A_{n,k}$ are defined in the following way: a transposition $i_1, ... i_{n+1}$ of numbers $1, ..., n+1$ is said to have a decrease at $l$ if $i_l > i_{l+1}$. Then $A_{n,k}$ is the number of transpositions of $n+1$ numbers, having $k$ decreases.

So $f_n = \sum_{k=0}^n A_{n,k} \alpha^n \beta^{n-k} {1 \over (n+1)!}$.

On the other hand, we have $g_2 = {1 \over 12} (\mu_1^2 + 4 \mu_2)^2$, $g_3 = - {1 \over 6^3} (\mu_1^2 + 4 \mu_2)^3$, so formula (\ref{exp}) gives the exponential of the formal group law $F_{\mu}$:
\begin{equation} \label{mu12}
f(u) = - 2 {\wp(u) -  {1 \over 12}(\mu_1^2 + 4 \mu_2) \over \wp'(u) - \mu_1 \wp(u) + {1 \over 12} \mu_1 (\mu_1^2 + 4 \mu_2)},
\end{equation}
where $\wp(u) = \wp(u; g_2, g_3)$. It implies that $f(u)$ gives a solution of the equation (\ref{Ricatti}).

\end{example}

\begin{rem} \label{rem61}
{\bf The $\wp$-function for $\Delta = 0$}.

We see that in the considered case $\Delta = g_2^3 - 27 g_3^2 = 0$.
Set $g_2 = {4 \over 3} \gamma^4$, $g_3 = - {8 \over 27} \gamma^6$.
In this case (see \cite{ref1})
\[
\sigma(z, g_2, g_3) = {1 \over 2 \gamma} e^{ - {1 \over 6} \gamma^2 z^2 } (e^{ \gamma z} - e^{ - \gamma z}),
\]
thus
\[
\wp(z, g_2, g_3) = - \zeta' = - {2 \over 3} \gamma^2 + \gamma^2 cth(\gamma z)^2.
\]

It follows from formula (\ref{mu12}) that in the considered case
\[
f(u) = {1 \over \gamma cth(\gamma z) + {1 \over 2} \mu_1} \text{ for } \gamma = {1 \over 2} \sqrt{\mu_1^2 + 4 \mu_2}.
\]

\end{rem}

\begin{cor} \label{cor1} In the case $\mu_2 = \mu_3 = \mu_4 = \mu_6 = 0$
we get
\[
F_{\mu}(t_1, t_2) = t_1 + t_2 - \mu_1 t_1 t_2,
\]
so $F_{\mu}$ is the multiplicative formal group.

The exponential satisfies the equation
\[
f(u)' = 1 - \mu_1 f(u).
\]
Using the condition $f(u) = u + ...$, we get
\begin{equation} \label{mu1}
\mu_1 f(u) = 1 - e^{- \mu_1 u}.
\end{equation}
On the other hand, (\ref{exp}) gives
\[
f(u) = - 2 {\wp(u; {1 \over 12} \mu_1^4, - {1 \over 6^3} \mu_1^6) -  {1 \over 12} \mu_1^2 \over \wp'(u; {1 \over 12} \mu_1^4, - {1 \over 6^3} \mu_1^6) - \mu_1 \wp(u; {1 \over 12} \mu_1^4, - {1 \over 6^3} \mu_1^6) + {1 \over 12} \mu_1^3}.
\]
Using remark \ref{rem61} we see that this formula for $f(u)$ conclides with (\ref{mu1}).

It follows from (\ref{mu1}) that
\[
f(u) =  u + \sum (-1)^{n} \mu_1^{n} {u^{n+1} \over (n+1)!}.
\]
Thus $B_f$ is generated by $f_n = {(-1)^{n} \mu_1^{n} \over (n+1)!}$, and
\[
B_f = \mathbb{Z}[b_n] / J
\]
where $J$ is an ideal generated by polynomials $(n+1)! b_n - 2^n b_1^{n}$.

\end{cor}

\begin{cor} \label{cor2} In the case $\mu_1 = \mu_3 = \mu_4 = \mu_6 = 0$
we get
\[
F_{\mu}(t_1, t_2) = { t_1 + t_2 \over 1 + \mu_2 t_1 t_2},
\]
which is a formal group coming from the addition formula for the hyperbolic tangent function.

The exponential satisfies the equation
\[
f(u)' = 1 - \mu_2 f(u)^2.
\]
Using the condition $f(u) = u + ...$, we get
\[
f(u) = {1 \over \sqrt{\mu_2}} th(\sqrt{\mu_2} u).
\]
Thus
\[
f(u) = \sum_{k=1}^{\infty} 2^{2k} (2^{2k} - 1) B_{2k} \mu_2^{ k - 1} {u^{2 k - 1} \over (2k)!},
\]
where $ B_{2k} $ denotes the Bernoulli numbers.
It follows that $B_f$ is generated by $f_{2 k - 2} = {2^{2k} (2^{2k} - 1) B_{2k} \mu_2^{k - 1} \over (2k)!}$ and we have $f_{2k-1} = 0$. Thus $f_{0} = 6 B_{2} = 1$, $f_{2} = 10 B_{4} \mu_2 = - {1 \over 3} \mu_2$.
\[
B_f = \mathbb{Z}[b_{2k} ] / J
\]
where $J$ is an ideal generated by the polynomials $(2k)! b_{2 k - 2} - 2^{2k} (2^{2k} -$~$1) B_{2k} (- 3 b_{2})^{k - 1}$.

\end{cor}

\begin{example} \label{ex6} Let $\mu_1 = \mu_2 = \mu_4 = \mu_6 = 0$.

In this case we obtain an elliptic curve in Tate coordinates (\ref{Ats})
\[
s =  t^3 + \mu_3 s^2, \quad \text{ with } \quad \Delta = - 27 \mu_3^4.
\]
The elliptic formal group law (\ref{123}) takes the form
\begin{equation} \label{F3}
F_{\mu}(t_1, t_2) =  {  (t_1 + t_2) - \mu_3 (t_1 + t_2) b - \mu_3 t_1 t_2 m \over (1 - \mu_3 b)^2}.
\end{equation}
Thus
\[
F_{\mu}(t_1, t_2) = (t_1 + t_2) - \mu_3 t_1 t_2 (2 t_1^2 + 3 t_1 t_2 + 2 t_2^2) + O(t^7).
\]
Though $\text{gcd}(2,3) = 1$, we obtain the homomorphism $\mathcal{A} \to \mathbb{Z}[\mu_3]$ classifying the formal group law (\ref{F3}) to be an epimorphism, that is there exists a set of generators $\{a_n\}$ in $\mathcal{A}$, such that $\phi(a_n) = 0$ for $n \ne 3$, $\phi(a_3) = \mu_3$.

Formula (\ref{geneq}) gives an equation on the exponential $f(u)$ of the formal group law $F_{\mu}$:
\begin{equation} \label{f3}
f'(u)^2 = - 4 \mu_3 f(u)^3 + 1.
\end{equation}
Therefore, we get
\[
f(u) = {- 1 \over \mu_3} \wp(u+c; 0, - \mu_3^2),
\]
where $c$ is given by the conditions
\[
\wp(c; 0, - \mu_3^2) = 0, \quad \wp'(c; 0, - \mu_3^2) = - \mu_3.
\]
On the other hand, we have $g_2 = 0$, $g_3 = - \mu_3^2$, so the function
\[
f(u) = - 2 { \wp(u; 0, - \mu_3^2) \over \wp'(u; 0, - \mu_3^2) - \mu_3}
\]
is the exponential of the formal group  $F_{\mu}$. It implies that
\[
{- 1 \over \mu_3} \wp(u+c; 0, - \mu_3^2) = - 2 { \wp(u; g_2, g_3) \over \wp'(u; g_2, g_3) - \mu_3}.
\]

\end{example}

\begin{example} \label{ex7}  Let $\mu_3^2 = - 3 \mu_6$, $2 \mu_3 \mu_4 = 3 \mu_1 \mu_6$, $\mu_4^2 = 3 \mu_2 \mu_6$, $\mu_6 \ne 0$.

In this case we obtain an elliptic curve in Tate coordinates (\ref{Ats})
\[
s =  t^3 + \mu_1 t s + \mu_2 t^2 s + \mu_3 s^2 + \mu_4 t s^2 + \mu_6 s^3.
\]
Formula (\ref{geneq}) gives an equation on the exponential $f(u)$ of the formal group law:
\begin{equation} \label{cube}
f'(u)^3 = ((1 - {1 \over 2} \mu_1 f(u))^3 - 3 \mu_3 f(u)^3)^2.
\end{equation}

We have $
g_2 = 0,
\quad
g_3 = - \mu_6 = {1 \over 3} \mu_3^2,
\quad
\Delta = g_2^3 - 27 g_3^2 = - 3 \mu_3^4$.

It follows from (\ref{exp}) that the solution of (\ref{cube}) is
\[
f(u) = - 2 { \wp(u; 0, {1 \over 3} \mu_3^2) \over \wp'(u, 0, {1 \over 3} \mu_3^2) - \mu_1 \wp(u; 0, {1 \over 3} \mu_3^2) - \mu_3}.
\]

\end{example}

\section{\bf Acknowledgments.}
The authors express their gratitude to I.~M.~Krichever and O.~M.~Mokhov for valuable comments during the discussion of this work.

\end{document}